\documentclass[aps,prb,twocolumn,nopacs,10pt,a4paper,nofootinbib,floatfix]{revtex4}

\usepackage[utf8]{inputenc}
\usepackage{graphicx,amsmath}
\usepackage{amssymb}
\usepackage{amsthm}
\usepackage{slashed}
\usepackage{braket}
\usepackage{color}
\usepackage{hyperref}
\usepackage{tikz}
\usetikzlibrary{plotmarks}
\usepackage{bm}
\allowdisplaybreaks[4]

\begin{document}
\title{Zero modes of the Kitaev chain with phase-gradients and longer range
couplings}
\author{Iman Mahyaeh}
\author{Eddy Ardonne}
\affiliation{Department of Physics, Stockholm University, SE-106 91 Stockholm, Sweden}
\date{\today}

\begin{abstract}
We present an analytical solution for the full spectrum of Kitaev's one-dimensional $p$-wave
superconductor with arbitrary hopping, pairing amplitude and chemical potential in the case of an open chain. We also discuss the structure of the zero-modes in the presence of both phase gradients and next nearest neighbor hopping and pairing terms. As observed by Sticlet et al., one feature of such models is that in a part of the phase diagram, zero-modes are present at one end of the system, while there are none on the other side. We explain the presence of this feature analytically, and show that it requires some fine-tuning of the parameters in the model. Thus as expected, these `one-sided' zero-modes are neither protected by topology, nor by symmetry.
\end{abstract}
\maketitle
\section{Introduction}

One of the characteristic features of many topological phases is the presence of gapless boundary
modes. The (fractional) quantum Hall states are a prime example
\cite{k80,tsg82,wen-adv}, and their boundary modes
provide strong evidence of the topological nature of these states. Another prime example is
the Kitaev chain, whose topological p-wave superconducting phase features so-called
`Majorana zero modes' at its edges.\cite{kitaev01}
Trying to establish the existence of the topological phase is often done by trying to
establish the presence of the boundary modes. This
has led to strong evidence for the topological phase in for instance
strongly spin-orbit coupled nano-wires
that are proximity coupled to an s-wave superconductor in the presence of a magnetic field
\cite{oreg10,lytchyn10,mourik12,deng12,das12},
or in chains of magnetic ad-atoms\cite{choy11,peintka13,nadj-perge13,nadj-perge14}.
It has been proposed that the zero energy Majorana bound states can be used as topologically
protected q-bits, for quantum information processing purposes \cite{kitaev03,alicea12}. By now, there exist various proposals to manipulate these q-bits, either in T-junction systems, in which the Majorana bound states can be braided explicitly\cite{alicea11}, or in Josephson coupled Kitaev chains, in which the coupling of the various chains allows operation on the q-bits\cite{aasen16}.

Despite the intense research on the Kitaev chain models, there are still interesting features that
deserve attention. In this paper, we look into one of them.
It was observed by Sticlet et al.\cite{bena13}, that
the zero-modes of Kitaev chains carrying a current, i.e.,
in the presence of a gradient in the phase of the order parameter,
have interesting properties. The most striking feature is that is it possible that
at one edge of the chain, there is pair of Majorana bound states (or better, one `ordinary'
Dirac zero mode), while there is no zero mode at the other end of the chain. Clearly, from a
topological point of view, this means that the chain is in a trivial phase, but it is nevertheless worthwhile
to investigate these zero-modes further. In this paper, we explain the presence of these zero-modes,
via an exact solution of the zero modes of an extended Kitaev chain,
i.e., in the presence of both complex and next nearest-neighbor hopping an pairing terms.
We show that it is necessary to fine tune the couplings in order that these
`one-sided Dirac modes' to exist, but under these fine-tuned conditions, they can only disappear
if the bulk gap closes signaling a phase transition or a crossover. Dropping the fine-tuning will gap out these zero modes immediately, leaving behind low-energy subgap modes. Apart from the analytical solution of the zero modes, we also present the solution of the full spectrum of the open Kitaev chain, for
real, but otherwise arbitrary parameters, which does not seem to have appeared in the literature before.

The outline of the paper is as follows. We start in Sec.~\ref{secLSM} by a brief review of the
Lieb-Schultz-Mattis method to solve open quadratic fermionic systems,
and focus on the case of complex couplings, which is essential for our purposes.
In Sec.~\ref{sec.XY in TF}, we provide the full solution of the open Kitaev chain, with real, but
otherwise arbitrary couplings. In Sec.~\ref{sec:long-range-complex}, we study the
effect of next nearest-neighbor and complex couplings.
Here, we focus entirely on the exact solutions for the
zero-modes, and start by considering the effects of next nearest-neighbor couplings
and complex pairings separately, before coming to the most interesting case, when both are present.  
In Sec.~\ref{sec:discussion}, we discuss the results of the paper. Some details are delegated to
the appendices.

\section{The Lieb-Schultz-Mattis method}  
\label{secLSM}

In this paper, we study the zero modes of Kitaev-like chains in the presence of longer range hopping and pairing terms, specifically next nearest neighbor (NNN) ones. In particular, we are interested in the case where these couplings are complex. To study these systems, we use the method has been introduced by Lieb, Schultz and Mattis (LSM)\cite{lsm61} who used it to solve the XY chain, for various types of boundary conditions. For a quadratic fermionic Hamiltonian with periodic boundary conditions (PBC), one diagonalizes the Hamiltonian by using a Fourier transformation, followed by a Bogoliubov transformation in the case of superconducting model.
Without translational invariance one can still perform a Bogoliubov like transformation directly in real space. It was this method that LSM used to find the spectrum of the open XY chain (after using a Jordan-Wigner transformation to transform the spin degrees of freedom to polarized fermions). 

In this section we review the LSM method and follow their notation for convenience. We consider two different cases. First, we look at the Hamiltonian with real couplings and recall how one can derive the spectrum of the model analytically. Second, for a general quadratic Hamiltonian with complex couplings we present the equations governing the zero mode solutions, which we use throughout the paper.

Following LSM\cite{lsm61}, we consider the general quadratic Hamiltonian of polarized fermions as follows,
\begin{equation}
\label{eq:LSM convention}
H= \sum_{i,j=1}^N c^{\dagger}_i A_{ij}c_j +\dfrac{1}{2}(c^{\dagger}_i B_{ij}c^{\dagger}_j +h.c. ),
\end{equation} 
in which $c_i$ is a fermion annihilation operator on site $i$, $A$ is a hermitian matrix, $B$ is an antisymmetric matrix and $N$ is the number of sites. Using a Bogoliubov like transformation, one can define new fermion operators, and diagonalize the Hamiltonian:
\begin{eqnarray}
\label{eq:eta- c fermion}
\eta_{\alpha}=\sum_{i=1} ^N(g_{\alpha,i}c_i +h_{\alpha,i}c^{\dagger}_i), \\
H=\sum_{\alpha=1}^N \Lambda_{\alpha}\eta^{\dagger}_{\alpha}\eta_{\alpha},
\end{eqnarray}
in which $\alpha$ labels the states and $g_{\alpha,i}$ and $h_{\alpha,i}$ are two functions, which are to be determined. This transformation is canonical, in the sense that new operators obey the fermionic anti-commutation relations, i.e. $\{\eta_{\alpha}, \eta ^{\dagger}_{\beta} \}=\delta_{\alpha\beta}$.  

Using the equation of motion, $[H,\eta_{\alpha}]=-\Lambda _{\alpha}\eta_{\alpha}$, one finds the equations for $g_{\alpha,i}$ and $h_{\alpha,i}$:
\begin{eqnarray}
\label{eq:g&h1}
h_{\alpha,i}B^*_{ij}-g_{\alpha,i}A_{ij}= -\Lambda_{\alpha} g_{\alpha,j},\\
\label{eq:g&h2}
 h_{\alpha,i}A^*_{ij}-g_{\alpha,i}B_{ij}=- \Lambda_{\alpha} h_{\alpha,j}.
\end{eqnarray}

In order to find the full spectrum of the Hamiltonian, we now consider the case for which $A$ and $B$ have real elements. In this case, one defines new variables as
\begin{eqnarray}
\phi_{\alpha,i}=g_{\alpha,i} + h_{\alpha,i}, \\
\psi_{\alpha,i}=g_{\alpha,i} - h_{\alpha,i} ,
\end{eqnarray}
which we combine into row vectors, $\phi_{\alpha}=(\phi_{\alpha,1},\dots,\phi_{\alpha,N})$ and $\psi_{\alpha}=(\psi_{\alpha,1},\dots,\psi_{\alpha,N})$. Summing and subtracting Eqs.\eqref{eq:g&h1} and \eqref{eq:g&h2} gives two coupled equations for $\phi_{\alpha}$ and $\psi_{\alpha}$,
\begin{eqnarray}
\label{eq:phi(A-B)}
\phi_{\alpha} (A-B)=\Lambda_{\alpha} \psi_{\alpha},\\
\label{eq:psi(A+B)}
\psi_{\alpha}(A+B)=\Lambda_{\alpha} \phi_{\alpha}.
\end{eqnarray}
We note that the matrices act from the right side on the vectors. By acting with $A+B$ on Eq.\eqref{eq:phi(A-B)} and $A-B$ on Eq.\eqref{eq:psi(A+B)} from the right, the equations decouple
\begin{eqnarray}
\label{eq:finding phi}
\phi_{\alpha} (A-B)(A+B)=\Lambda_{\alpha}^2 \phi_{\alpha},\\
\label{eq:finding psi}
\psi_{\alpha} (A+B)(A-B)=\Lambda_{\alpha}^2 \psi_{\alpha}.
\end{eqnarray}
To find all the eigenvalues $\Lambda_\alpha$ and states $\eta _{\alpha}$, one solves these
two decoupled equations for $\phi_{\alpha}$ and $\psi_{\alpha}$. We explain how to do this
in more detail in the next section for the {\em open} Kitaev chain\cite{kitaev01} with real, but
otherwise generic parameters. 

It is well-known that fermionic systems can host Majorana zero modes on the edges of the system, which
signals that the system is in a topological phase. In this paper, we study the zero modes of Hamiltonians with complex parameters, so we now allow the matrices $A$ and $B$ to be complex again. To distinguish a Majorana mode from the ordinary modes, we use stared labels, such as $\alpha^*$. The Majorana modes satisfy $\eta_{\alpha^*}=\eta^{\dagger}_{\alpha^*}$.
For a finite system, the energy of a Majorana mode is exponentially small in the system size; for instance in the case where we have a system with $N$ sites the energy scales as $\Lambda_{\alpha^*} \sim e^{-\kappa N}$ with $\kappa >0$ \cite{lsm61,kitaev01}. Hence we are interested in finding general equations which allows one to find the corresponding states with zero energy, i.e. $\Lambda_{\alpha^*}=0$, in the thermodynamic limit.

We thus search for a Majorana solution of Eqs.\eqref{eq:g&h1} and \eqref{eq:g&h2} with zero energy. Setting $h_{\alpha^*,i}=g^*_{\alpha^*,i}$ in Eqs. \eqref{eq:g&h1} and \eqref{eq:g&h2} gives:
\begin{eqnarray}
g^*_{\alpha^*,i}B^*_{ij}=g_{\alpha^*,i}A_{ij},\\
 g^*_{\alpha^*,i}A^*_{ij}=g_{\alpha^*,i}B_{ij},
\end{eqnarray}
By summing and subtracting these equations we get,
\begin{eqnarray}
\label{eq:LSMreal}
\mathrm{Re}[g(A-B)]=0,\\
\label{eq:LSMimag}
\mathrm{Im}[g(A+B)]=0.
\end{eqnarray}  
We use these equations to explore the wave functions ($g_{\alpha^*,i}$) of the zero modes in different cases in the following sections.

Before closing this section, we write the $\eta$ operators in terms of Majorana operators for future reference. Using $\phi$ and $\psi$ as defined above and defining Majorana operators as $\gamma_{A,j}=c^{\dagger}_j+c_j$ and $\gamma_{B,j}=i (c^{\dagger}_j-c_j)$, we write the fermion annihilation operator as follows
\begin{equation}
\label{eq:etapsiphi}
\eta_{\alpha}=\sum_{j=1} ^N[\dfrac{\phi _{\alpha,j}}{2}\gamma_{A,j} +i\dfrac{\psi _{\alpha,j}}{2}\gamma_{B,j} ].
\end{equation}
The algebra of Majorana operators can be calculated from the canonical anti-commutation relations of the $c$ operators,
\begin{equation}
\label{eq: Majorana commutation}
\{\gamma_{r,i},\gamma_{r',j}\}=2\delta_{rr'}\delta_{ij}.
\end{equation}
Specifically, for the zero mode solution we can write the corresponding fermionic operator as follows:
\begin{equation}
\label{eq:majorana}
\eta_{\alpha^*}=\sum_{j=1} ^N(\mathrm{Re}[g_{\alpha^*,i}] \gamma_{A,j} -\mathrm{Im}[g_{\alpha^*,i}]\gamma_{B,j} ).
\end{equation}

\section{The spectrum of the open Kitaev chain}
\label{sec.XY in TF}
In this section, we use the method of LSM to find the full spectrum of the Kitaev chain \cite{kitaev01},
for an open chain, with real parameters, in particular we consider
\begin{equation}
H=\dfrac{1}{2}\sum_{j=1}^{N-1}(c^{\dagger}_jc_{j+1}+\Delta c^{\dagger}_jc^{\dagger}_{j+1}+h.c.) -\mu \sum_{j=1}^{N}(c^{\dagger}_jc_{j}-\dfrac{1}{2}) \ .
\end{equation}
Here, $\mu$ denotes the chemical potential, $\Delta$ the strength of the pairing term, and we
chose the hopping parameter $t = -1$
\footnote{The sign of $t$ is irrelevant for the spectrum, but we set $t=-1$, because of the simpler
relation with the associated $XY$ model as studied in\cite{lsm61}.}.

Despite the fact that this model has been studied thoroughly, these results do not seem to have appeared
in the literature, and we will use it to set the notation. Because we are interested in the zero-modes
of more generic situations in the remainder of the paper, we also quickly review the nature of the zero-modes.
These latter results are not new, but appeared in \cite{lsm61,epw70,p70} and for generic parameters recently
in \cite{hegde16,ortiz17}.

It is well known\cite{kitaev01} that the Kitaev chain is in a topological phase for
$|\mu| < |t|$ and $\Delta \neq 0$.  A profound feature of topological phase
is the presence of a Majorana zero modes, that are exponentially localized near the edges of the
system. In addition, the energy associated with this zero mode is exponentially small in the system
size. 

To set the scene, we follow Kitaev to show the presence of Majorana zero modes, by considering
the special case of $\Delta=1$ and $\mu =0$. In this case, the Hamiltonian in terms of
Majorana operators becomes,
\begin{equation}
\label{eq:kitaevsimple}
H=-\dfrac{i}{2}\sum_{j=1}^{N-1}\gamma_{B,j}\gamma_{A,j+1}.
\end{equation}

In this Hamiltonian, $\gamma_{A,1}$ and $\gamma_{B,N}$ are absent and therefore commute with it. So one can form a non-local fermionic state, $f_0= \frac{1}{2}(\gamma_{A,1} + i \gamma_{B,N})$. The presence of this non-local fermionic mode is the characteristic feature of the topological phase of the Kitaev chain. For $\Delta=-1$, the unpaired Majorana operators would be $\gamma_{B,1}$ and $\gamma_{A,N}$, owing to the p-wave nature of pairing.  

We leave this fine tuned point and consider arbitrary $\Delta$, but keep $\mu=0$ for the moment. This corresponds to the XY model, which was solved exactly by LSM for $|\Delta|<1$, that is, the full spectrum including the wave functions were found \cite{lsm61}. For an open chain, there is a state with an exponentially small energy as a function of the system size. The wavefunction of this state is exponentially localized on the edges, namely $\phi_n \sim \big(\frac{1-|\Delta|}{1+|\Delta|}\big)^n$ where $n$ denotes the position of the site measured from the left side of the chain. The associated $\psi_n$ is localized on the right edge. Another fine tuned point that was studied previously corresponds to the transverse field Ising model(TFIM), that is $t=-1$, $\Delta=\pm1$ but arbitrary $\mu$.
Pfeuty showed that this model has a Majorana zero mode if $|\mu|<1$. The associated wave function
takes the form $\phi_n \sim |\mu|^n$ and is localized on the left edge of the system\cite{epw70,p70}.

To find the Majorana zero modes for the general case, it is advantageous to first consider the
model with periodic boundary conditions. That is, we need to consider the hopping and
pairing terms for the last site as well. We denoted the periodic Hamiltonian by
$H_{PBC}=H + H_N$ where:
\begin{equation}
H_N=\dfrac{1}{2}(c^{\dagger}_Nc_{1}+\Delta c^{\dagger}_Nc^{\dagger}_{1}+h.c.),
\end{equation}

The solution of the periodic model is well known, and obtained by using a plane-wave
ansatz for the wave functions (i.e., by Fourier-transforming the model).
Using the method outlined in the previous section, we start by solving
Eqs.~\eqref{eq:finding phi} and \eqref{eq:finding psi} to find the spectrum. Since $\phi$ and $\psi$ are
related via Eqs.~\eqref{eq:phi(A-B)} and \eqref{eq:psi(A+B)}, we focus on $\phi$.
Writing Eq.~\eqref{eq:finding phi} gives us one recursion relation:
\begin{align}
\label{eq: bulk equation}
&(1-\Delta^2)\phi_{\alpha,n-2} - 4\mu \phi_{\alpha,n-1} + [4\mu^2 +2(1+\Delta^2)]\phi_{\alpha,n} \nonumber \\
+&(1-\Delta^2)\phi_{\alpha,n+2}-4\mu \phi_{\alpha,n+1} 
=4\Lambda^2_{\alpha}\phi_{\alpha,n},
\end{align} 
where $n$ denotes the sites and runs from $1$ to $N$. Upon
setting $\phi_{k,n} \sim e^{i k n}$, where we use the momentum $k$ as a label, one
finds the eigenvalues:
\begin{equation}
\label{eq:dispersion}
\Lambda_k^2=(\mu - \cos k)^2+\Delta^2 \sin^2k,\hspace{0.5cm} k=\frac{2 \pi m}{N},
\end{equation} 
where $m$ runs over $0$ to $N-1$. If one considers anti-periodic boundary conditions, the dispersion is the same though the allowed values of $k$ change to $k=\frac{2\pi}{N}(m+\frac{1}{2})$. To study open chains, we make use the functional form of the dispersion. In addition, by using the LSM method for open chains, we naturally have to consider both the sectors with even and odd number of particles.

We now consider the full spectrum of the open chain. Here, we merely give the results, and refer to App.~\ref{sec:Appendix GKC}, where the details of calculation are presented.

For the open chain we find the same recursion relation in the bulk which is valid for $3\leq n\leq N-2$. However, we also have four boundary equations which should be treated separately (see App.~\ref{sec:Appendix GKC}). We start by dealing with the bulk equations, using the method of LSM. That
is, we use the same `function' for the eigenvalues, though with a generic parameter $\alpha$ instead of
the momentum $k$. To find the allowed values for the parameter $\alpha$, one uses the `boundary equations'. Hence we parametrize the eigenvalues as:
\begin{equation}
\label{eq:eigenGKC}
\Lambda^2_{\alpha}=(\mu - \cos \alpha)^2+\Delta^2 \sin^2\alpha,
\end{equation}    
and $\alpha$ is the label for the state. For the states, we use a power law ansatz, $\phi_{\alpha,n}\sim x_{\alpha}^n$, and we find four solutions,
$x_{\alpha}=e^{\pm i \alpha}$ and $x_{\alpha}=e^{\pm i \beta}$ where
\begin{equation}
\label{eq:alphabeta1}
\cos \alpha + \cos \beta=\dfrac{2 \mu}{1- \Delta^2}.
\end{equation}
Note that $\alpha$ and $\beta$ are not necessarily real, but the way we parametrize $x_{\alpha}$ turns out to be convenient.
As described in the App.~\ref{sec:Appendix GKC}, the relevant linear combination that one
uses to find a solution for the boundary equations is:
\begin{align}
\label{eq: ansatz}
\phi_{\alpha,n}&=A_1 \Big\{ \sin[(N+1)\beta] \sin(n\alpha) \nonumber\\
&- \sin[(N+1)\alpha] \sin(n \beta) \Big\}\nonumber \\
& + A_2 \Big\{ \sin[(N+1)\beta] \sin[(N+1-n)\alpha] \nonumber \\
&- \sin[(N+1)\alpha] \sin[(N+1-n) \beta] \Big\} .
\end{align}
in which $A_1$ and $A_2$ are constants that are related via Eq.~\eqref{eq:appA A1 A2}. The boundary equations give rise to another constraint on $\alpha$ and $\beta$. This constraint can be shown
to take the following form
\begin{align}
\label{eq:alphabeta2}
&\sin^2\alpha +\sin^2\beta + \dfrac{1}{\Delta^2}(\cos\beta-\cos\alpha)^2\nonumber \\
&-2\dfrac{\sin \alpha \sin \beta}{\sin[(N+1)\alpha]\sin[(N+1)\beta]}\nonumber \\
&\times\{1-\cos[(N+1)\alpha]\cos[(N+1)\beta]\}=0.
\end{align}
To obtain the full solution of the model, one needs to solve Eqs.~\eqref{eq:alphabeta1} and
\eqref{eq:alphabeta2} simultaneously. Though this can not be done analytically, it is straightforward
to obtain the solutions numerically. Thus, we have characterized all the eigenvalues and
eigenvectors $\phi_{\alpha,n}$ and by using Eq.~\eqref{eq:finding phi}, one finds $\psi_{\alpha,n}$. 

Now we want to study these solutions and see when this model has a Majorana solution and what the corresponding wavefunction is. To find such solutions, we consider thermodynamic limit, i.e. $N \rightarrow \infty$, which makes the calculations easier.
 
We first mention that $\Delta$ can always set to be positive. One way to see this is by considering the transformation under which $c_j$ maps to $e^{i \frac{\pi}{2}} c_j$. This transformation changes neither the hopping nor chemical potential term, but $\Delta$ changes to $-\Delta$. In addition solutions for $\mu < 0$ can be constructed from the solutions for $\mu > 0$. One can take the solution for $\mu>0$, say $(\alpha,\beta)=(r,s)$. Now consider $(\alpha,\beta) = (r+\pi,s+\pi)$. This change gives a minus sign for the LHS of Eq.~\eqref{eq:alphabeta1} as required, however it leaves Eq.~\eqref{eq:alphabeta2} unchanged. Therefore, we restrict ourselves to $\Delta,\mu > 0$.

First we look at the solutions for large values of $\mu$.
In this case one can see that Eqs.~\eqref{eq:alphabeta1} and \eqref{eq:alphabeta2} have $N$ distinct real solutions for $\alpha$, where we restrict $\alpha$ to lie in the range $0 < \alpha \leq \pi$ ($\alpha = 0$ gives $\phi_n = 0$; for more details, see 
App.~\ref{sec:Appendix GKC}).
However by decreasing chemical potential solution with the smallest value of
$\alpha$ will `disappear'. It is well known that
for $\mu<1$ one real solution is lost in the thermodynamic limit. For a finite chain this happens for $\mu < 1 + O(\frac{1}{N})$ where $O(\frac{1}{N})$ is a finite size correction. 
Thus, for $\mu < 1$ one must find an additional, complex solution.
To find this solution, we consider three different cases.

\textit{1) $\Delta < 1$ and $\sqrt{1- \Delta^2}<\mu<1$}:  In the thermodynamic limit one can check that the following solution satisfies Eqs.\eqref{eq:alphabeta1} and \eqref{eq:alphabeta2},
\begin{eqnarray}
&\alpha^*=i(\dfrac{1}{\xi_1}+\dfrac{1}{\xi_2}),&\ \beta^*= i(\dfrac{1}{\xi_1}-\dfrac{1}{\xi_2}),\\
\label{eq:MZM decay}
&\cosh \dfrac{1}{\xi_1} = \dfrac{1}{\sqrt{1-\Delta^2}},& \ \cosh \dfrac{1}{\xi_2} = \dfrac{\mu}{\sqrt{1-\Delta^2}}.
\end{eqnarray} 
Furthermore it is straightforward to check that Eq.\eqref{eq:eigenGKC} gives $\Lambda_{\alpha^*}=0$, hence the solution is indeed a zero mode. The wave function $\phi_{\alpha^*,n}$ for this zero mode is
\begin{equation}
\label{eq:phi decay}
\phi_{\alpha^*,n}=C e^{-\dfrac{n}{\xi_1}} \sinh(\dfrac{n}{\xi_2}),
\end{equation}
where $C$ is a normalization constant. Moreover, it can be shown that based on structure of $A-B$ and $A+B$ matrices, one has $\psi_{\alpha^*,n}=\phi_{\alpha^*,N+1-n}$. From the fact that $\xi_1< \xi_2$, it follows that $\phi_{\alpha^*}$ is localized on the left edge while $\psi_{\alpha^*}$ is localized on right edge of the system. Hence we found two Majorana operators, that are located on the edges of the system, and the associated wavefunctions decay exponentially.

\textit{2) $\Delta < 1$ and $\mu<\sqrt{1- \Delta^2}$}:
In this range, one needs to use a different parametrization if one wants to use real
parameters, as is evident from Eq.\eqref{eq:MZM decay}.
This parametrization reads
\begin{eqnarray}
\label{eq:oscdecay}
&\alpha^* =q+i\dfrac{1}{\xi},& \ \beta^*= q-i\dfrac{1}{\xi},\\
&\cos q = \dfrac{\mu}{\sqrt{1-\Delta^2}},& \ \cosh \dfrac{1}{\xi} = \dfrac{1}{\sqrt{1-\Delta^2}}. 
\end{eqnarray}
Basically we changed one of the characteristic length scales to become a wave vector.   
As in the previous case, this solution is indeed a zero mode, i.e. $\Lambda_{\alpha^*}=0$, whose wavefunction is given by:
\begin{equation}
\label{eq:phi oscdecay}
\phi_{\alpha^*,n}=C e^{-\dfrac{n}{\xi}} \sin(nq).
\end{equation}
This result indicates that $\phi$ ($\psi$) is localized on the left (right) edge with an oscillatory decaying wave function. We should point out that this result was obtained earlier by \cite{hegde16}.
In addition, it was observed that the correlation functions in the model with PBC are oscillatory in the same regime, i.e., for $\mu<\sqrt{1- \Delta^2}$ with $\Delta < 1$, see for instance
Refs.~\onlinecite{barouch71,suzuki71,book:henkel}.

\textit{3) $\Delta > 1$}: For this regime $\sqrt{1-\Delta^2}$ is imaginary, hence the previous solutions are not applicable. The new root can be written as
\begin{eqnarray}
\label{eq:Delta >1}
&\alpha^*=i(\dfrac{1}{\xi_1}-\dfrac{1}{\xi_2}),&\ \beta^*=\pi + i(\dfrac{1}{\xi_1}+\dfrac{1}{\xi_2}),\\
&\sinh \dfrac{1}{\xi_1} = \dfrac{1}{\sqrt{\Delta^2-1}},& \  \sinh \dfrac{1}{\xi_2} = \dfrac{\mu}{\sqrt{\Delta^2-1}}. 
\end{eqnarray}  
One can check that this solution represents a zero mode with the wave function:
\begin{equation}
\phi_{\alpha^*,n}=C e^{-\dfrac{n}{\xi_1}} \times \left\{
\begin{array}{rl}
 \cosh(\dfrac{n}{\xi_2}) & \text{if } n \ \text{is odd} ,\\
 \sinh(\dfrac{n}{\xi_2}) & \text{if } n \ \text {is even}.
\end{array} \right.
\end{equation}
For this solution $\xi_1 < \xi_2$ since $\mu<1$ and this guarantees that $\phi$ ($\psi$) is  localized on the left (right) edge.

\section{Zero-modes for next nearest-neighbour and complex couplings}
\label{sec:long-range-complex}

In this section we study the zero modes in the presence complex hopping and pairing terms, both
in the case with nearest neighbor hopping and pairing terms, as well as 
next-nearest neighbor (NNN) hopping and paring terms. The complex amplitudes model the
presence of a phase gradient in the system.

In their fermionic incarnation, these generalized Kitaev models were studied
in Refs.~\onlinecite{niu12,bena13,luca17,sen13}.
In the language of spin models, adding NNN terms gives rise to so-called (one-dimensional) cluster models%
\cite{doherty09,niu12,ohta15,lahtinen15,ohta16}, but we concentrate on the fermionic version of these models.

An important feature of these models is the possibility of having more than one zero modes at
each end, which is possible due to the presence of longer range terms.
This can also be understood in terms of the classification of topological insulators and superconductors\cite{kitaev09,ryu10}. The Kitaev chain with real coupling constants belongs to class BDI for which the different topological phases can be labeled by the elements of $\mathbb{Z}$, in the absence of interactions. Adding interaction changes this picture such that new classification is given by $\mathbb{Z}_8$ instead\cite{fidkowski10}. In the case with only nearest neighbor hopping and pairing terms, the model describes phases with at most one Majorana mode at each end of the system.
However by adding NNN terms one finds phases with two Majorana modes at each end. This means that there would be two distinct topological phases with one and two zero modes solutions (in addition to the trivial phase, which does not have a zero mode). 

Proposals for using the non-local fermionic state as a qubit, requires the ability to move Majorana edge states and even to do braiding. One proposal to achieve this is by inducing a phase gradient in the superconductor order parameter, i.e $\Delta_j =\Delta e^{i \theta_j}$, with non-uniform $\theta _j$\cite{romito12}. Having a complex superconductor order parameter breaks the time reversal symmetry in which case the model belongs to class D. For class D, we have the $\mathbb{Z}_2$ classification which means that the system could be either in the topological phase with at most one Majorana zero mode at each end, or in the trivial phase. Surprisingly, Sticlet et al. showed that NNN terms with a phase gradient can exhibit an exponentially localized fermionic zero mode on just one edge\cite{bena13}. Such a phase, though it is not topologically protected, has local zero modes. In Ref.~\onlinecite{bena13} these models were investigated numerically. Here we present an analytical solution and study the zero-modes in detail. We first review the Kitaev chain with NNN terms. After that, we study the effect of a constant phase gradient in the Kitaev chain. Finally, we combine the two complications and consider NNN terms and a phase gradient simultaneously. 

\subsection{Next nearest-neighbor couplings}
In this section we consider the Kitaev chain and add NNN hopping and pairing terms. We start with the case for which all the parameters are real, hence the Hamiltonian belongs to class BDI. Setting $\mu =1$, the problem has four energy scales, corresponding to the two hopping and two pairing amplitudes. To simplify the calculation we set the NN hopping and pairing terms equal to each other and we do the same for the NNN terms. Sticlet et al. studied this model under the same assumptions\cite{bena13}. We consider the model with arbitrary complex parameters in Sec.~\ref{sec:general parameters }.

\begin{figure}[th] 
\includegraphics[width=.9\columnwidth]{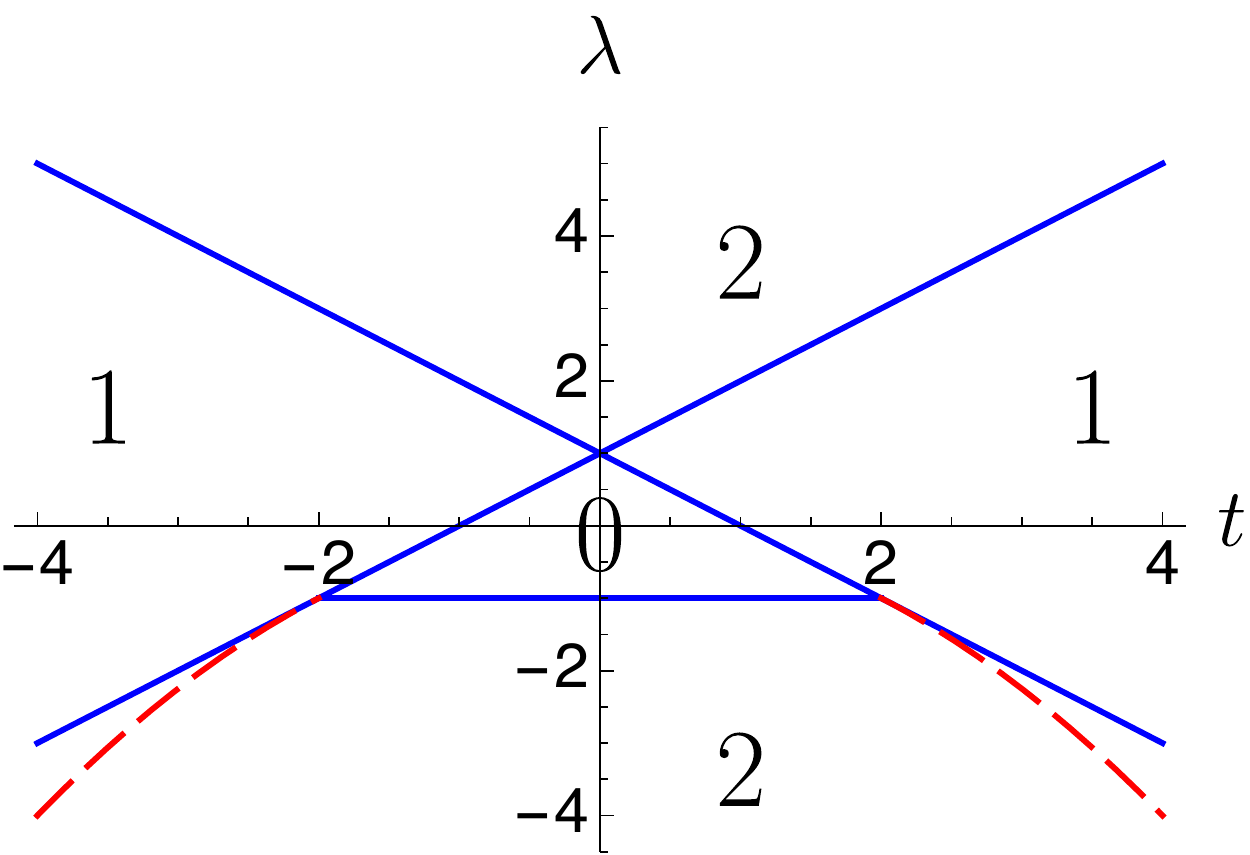}
\caption{Phase diagram for Kitaev chain with NNN terms for $\mu=1$ (Eq.~\eqref{eq:HKitnnn}). The numbers in the plot show the number of Majorana zero modes at each end of the chain. The solid lines represent phase boundaries. The line $\lambda=-1$ together with the dashed lines forms the boundary of the region in which $| x_\pm |=\frac{1}{|\lambda|}$. In this region the correlation length only depends on the NNN coupling.}  
\label{fig: PDNNN}
\end{figure}

Thus, the Hamiltonian reads,
\begin{align}
H&=\dfrac{t}{2}\sum_{j=1}^{N-1}(c^{\dagger}_jc_{j+1}+c^{\dagger}_jc^{\dagger}_{j+1}+h.c.) -\mu \sum_{j=1}^{N}(c^{\dagger}_jc_{j}-\dfrac{1}{2}) \nonumber \\
& + \dfrac{\lambda}{2}\sum_{j=1}^{N-2}(c^{\dagger}_jc_{j+2}+c^{\dagger}_jc^{\dagger}_{j+2}+h.c.) 
\label{eq:HKitnnn}
\end{align}
where $\lambda$ is the NNN hopping and pairing amplitude. To obtain the phase diagram, we first consider the model with periodic boundary conditions\cite{niu12,bena13}. We do a
Fourier transformation, $c_j=\dfrac{1}{\sqrt{N}}\sum_k e^{i k j}c_k$,
and define $\Psi_k=(c_k,c^{\dagger}_{-k})^\mathrm{T}$ to write the hamiltonian as
\begin{align}
&H= \dfrac{1}{2}\sum_k \Psi^{\dagger}_k \mathcal{H}_k \Psi_k, \nonumber \\
\mathcal{H}_k&= [-t  \sin(k) - \lambda  \sin(2k)]\tau^y \nonumber \\
&+[t  \cos(k) + \lambda  \cos(2k)-\mu]\tau^z,
\end{align}
where the $\tau^{\alpha}$ are Pauli matrices that act in the Nambu space $\Psi_k$. The Hamiltonian can be written as $\mathcal{H}_k=\mathbf{h}(k)\cdot \bm{\tau}$. One can find the phase diagram by calculating the winding number for $\mathbf{h} (k)$\cite{ryu10,bena13} or by looking at gap closing lines\cite{niu12}. The phase diagram is presented in Fig.~\ref{fig: PDNNN}. The gap closes along the lines $\lambda=\mu +t$, $\lambda=\mu -t$ and $\lambda =-\mu$ for $|t|<2|\mu|$. Note that in the figure we used $\mu=1$.

Before looking at the zero mode solution(s) of an open chain, we first consider some limiting cases to understand the phase diagram. For very small $|t|,|\lambda| \ll |\mu|$ we get the trivial phase. The $"0"$ in Fig.~\ref{fig: PDNNN} indicates that there are no Majorana zero modes in this part of the phase diagram. Outside of the trivial region on the vertical axis where $t=0$ we have two decoupled Kitaev chains, hence there are two zero modes at each end. For a fixed $\lambda$, adding NN terms couples these two chains. The two zero modes survive until the gap closes, thereafter there will only be one zero-mode at each end. The horizontal axis with $\lambda=0$ (i.e., the original Kitaev chain) belongs to this later region which is indicated by "$1$" in the Fig.~\ref{fig: PDNNN}.

To find the wave functions of the zero modes, we use Eqs.~\eqref{eq:phi(A-B)} and \eqref{eq:psi(A+B)} with $\Lambda_{\alpha}=0$. From Eq.~\eqref{eq:etapsiphi} we see that if $\eta_\alpha$ is a Majorana mode (i.e., $\eta_\alpha^\dagger = \eta_\alpha$), $\psi$ has to be purely imaginary. So for convenience we define $\psi= i \tilde{\psi}$ and we get $g=\frac{1}{2}(\phi +i \tilde{\psi})$. We use this convention from now on. We obtain the following
`bulk' equations
\begin{align}
-\mu \phi_{n} + t \phi_{n+1} + \lambda \phi_{n+2} &= 0\\
\lambda \tilde{\psi}_{n-2} + t \tilde{\psi}_{n-1} -\mu \tilde{\psi}_{n} &= 0 \ .
\end{align}
The `boundary' equations are $-\mu \phi_{N-1} + t \phi_{N} = 0$, $-\mu \phi_{N} = 0$,
$-\mu \tilde{\psi}_1 = 0$ and $t \tilde{\psi}_1 - \mu \tilde{\psi}_2 = 0$.
So we can use the ansatz $\phi_n \sim x_0 ^n$ and $\tilde{\psi}_n \sim x_0 ^{N-n+1}$, which
gives the result
\begin{align}
\label{eq:solution NNN}
&\phi_n= L_+ x^n_{0,+} + L_- x^n_{0,-} ,\nonumber\\
& \tilde{\psi}_n= R_+ x^{N-n+1}_{0,+} + R_- x^{N-n+1}_{0,-},\nonumber \\
&x_{0,\pm}=\dfrac{-t  \pm \sqrt{t^2+4\lambda \mu}}{2\lambda },
\end{align}
where $L_{\pm}$ and $R_{\pm}$ are real normalization constants (the subscript $"0"$ in the length scales indicates that we deal with a zero phase gradient). 

We can extract the phase diagram from this result\cite{niu12} and we set $\mu =1$ to be able to compare with Fig.~\ref{fig: PDNNN}. For regions where $\lambda > 1 +|t|$ or both $\lambda< 1 - |t|$ and $\lambda<-1$ (corresponding to the upper and lower regions of Fig.~\ref{fig: PDNNN}), one can see that $|x_{0,\pm}|<1$. This means that in these regions that are indicated by $"2"$ the system has two independent zero mode solutions. In the right part of the phase diagram where $1-t<\lambda<1+t$, there exists only one zero mode since $|x_{0,+}|<1$ and $|x_{0,-}|>1$. If $1+t<\lambda<1-t$ we also have one zero mode, however, in this case $|x_{0,-}|<1$ and $|x_{0,+}|>1$.
We note that in these regions, the boundary equations are also satisfied in the large $N$ limit.

It is also interesting to note that for $t^2+4\lambda>0$ the roots are real. Still they could be negative in some regions which gives rise to an oscillatory behavior of the wave functions, which are then proportional to $(-1)^n$. For $t^2+4\lambda<0$ the roots become complex. The red, dashed lines in Fig.~\ref{fig: PDNNN} specify the upper boundaries of this region (in the case $\lambda < -1$).
In this case $|x_{\pm}|=\frac{1}{\sqrt{|\lambda|}}$ which gives us the the criterion $\lambda < -1$ in order to have a zero mode (in the region $t^2+4\lambda<0$).
In this part of the phase diagram the correlation length only depends on $\lambda$, while the NN
coupling $t$ only affects the oscillatory part of the wave function.

Before moving to the case with both NNN terms as well as with a phase gradient, we first study the
Kitaev chain with just a constant phase gradient.

\subsection{Phase gradient in the order parameter}
\label{sec Pfeutyphase} 
In this subsection, we consider the Kitaev chain, but with a phase gradient in the
superconducting order parameter.
In the case of a superconductor with a super current, the pairing term has a site dependent phase
$\Delta_j=e^{i j \nabla \theta}$ where $\nabla \theta$ is the constant phase gradient, while
$j$ indicates the position of the site. In this case, the Hamiltonian reads
\begin{equation}
H=\dfrac{1}{2}\sum_{j=1}^{N-1}(c^{\dagger}_jc_{j+1}+e^{ i j \nabla \theta }c^{\dagger}_jc^{\dagger}_{j+1}+h.c.) -\mu \sum_{j=1}^{N}(c^{\dagger}_jc_{j}-\dfrac{1}{2}).
\end{equation} 
This Hamiltonian belongs to class D. As we indicated above, the topological phases are labeled by elements of $\mathbb{Z}_2$, which means that the system could be in the topological phase with one Majorana zero mode at each end. Changing the gauge, we transform the fermionic operators as $c_j\rightarrow e^{i j \frac{\nabla \theta}{2}} c_j$. This transformation gives us site-independent couplings, but now also the hopping parameter has become complex. The transformed Hamiltonian is
\begin{align}
H&=\dfrac{1}{2}\sum_{j=1}^{N-1}(e^{i \frac{\nabla \theta}{2}}c^{\dagger}_j c_{j+1}+e^{- i \frac{\nabla \theta}{2}}c^{\dagger}_j c^{\dagger}_{j+1}+h.c.) \nonumber \\
& -\mu \sum_{j=1}^{N}(c^{\dagger}_j c_{j}-\dfrac{1}{2}).
\end{align}
To find a zero mode solution we use Eqs.~\eqref{eq:LSMreal} and \eqref{eq:LSMimag}. The details of the solution for the Majorana operator are given in App.~\ref{sec:App Kitaev+phase}. The left Majorana solution is 
 \begin{equation}
 \label{eq:Pfeuty + phi left}
 \gamma_L=L \sum_{n=1}^{N}\Big[\dfrac{\mu}{\cos(\frac{\nabla \theta}{2})}\Big]^n \gamma_{A,n},
 \end{equation}
 where $L$ is the normalization constant to make $\gamma_L^2=1$ and the sum is over the sites. This Majorana mode is located at the left side of the system, and is a solution in the large $N$ limit.
The right Majorana mode is more complicated, 
\begin{align}
\label{eq:Pfeuty + phi right}
\gamma_R=&R  \sum_{n=1}^{N} \Big[\frac{\mu}{\cos(\frac{\nabla \theta}{2})}\Big]^{N-n+1} \times \nonumber \\
&[\sin(\frac{\nabla \theta}{2})\gamma_{A,n}+\cos(\frac{\nabla \theta}{2})\gamma_{B,n}],
\end{align}
where $R$ is the normalization constant to make $\gamma_R^2=1$ and the sum is over the sites. Using the Majorana modes $\gamma_L$ and $\gamma_R$, one can
construct one fermionic mode $f_0 = 1/2 (\gamma_L + i \gamma_R)$ as usual. 
We note that to have a localized zero mode we have the criteria $|\mu|< |\cos(\frac{\nabla \theta}{2})|$. This means that turing on the phase gradient shrinks the topological region. Second, we see that the left Majorana consists only of $\gamma_A$ Majorana operators (recall the definition above Eq.~\eqref{eq:etapsiphi}), however, the right one involves both $\gamma_B$'s as well as $\gamma_A$'s. In the case that $\nabla \theta=0$
the left Majorana mode only involves $\gamma_A$ operators and the right Majorana modes only $\gamma_B$ operators. This feature of the solution comes from the fact that for real $A$ and $B$ matrices (see Eqs.~\eqref{eq:LSMreal} and \eqref{eq:LSMimag}), the equations governing $\phi$ and $\tilde{\psi}$ are decoupled - recall that $g=\frac{1}{2}(\phi +i \tilde{\psi})$. Adding the phase gradient makes these matrices complex, hence the equations become coupled and the solutions become more complicated. The direction of the phase gradient shows itself in the elements of the $A$ and $B$ matrices and gives rise to this asymmetry; the "left-right" symmetry is broken explicitly. 

In the next section we add NNN terms to the current problem\cite{bena13}. The results presented in the current subsection are useful to understand zero mode solution(s) when one adds the NNN terms.

\subsection{Next nearest neighbor terms along with a phase gradient in the order parameter}
We now consider NNN terms in the presence of a constant phase gradient. Again we set the hopping and pairing amplitudes equal to each other for both the nearest neighbors and NNN terms. Following Sticlet et al \cite{bena13}, the Hamiltonian reads,
\begin{align}
\label{eq:hamil phase gradient with NNN}
H&=\dfrac{t}{2}\sum_{j=1}^{N-1}(c^{\dagger}_jc_{j+1}+e^{ ij \nabla \theta }c^{\dagger}_jc^{\dagger}_{j+1}+h.c.)  \nonumber \\
& + \dfrac{\lambda}{2}\sum_{j=1}^{N-2}(c^{\dagger}_jc_{j+2}+e^{ i j \nabla \theta}c^{\dagger}_jc^{\dagger}_{j+2}+h.c.) \nonumber \\
&-\mu \sum_{j=1}^{N}(c^{\dagger}_jc_{j}-\dfrac{1}{2}),
\end{align}
where we assumed the same phase dependence for the nearest neighbor and NNN pairing terms, with the same phase for both terms involving the first site.
As we mentioned above, for $\nabla \theta=0$ this model has a trivial phase without any zero modes and two topological phases that hosts one or two Majorana zero modes respectively (see Fig.~\ref{fig: PDNNN}). For $\nabla \theta \neq 0$, the model belongs to class D. This means that, contrary to $\nabla \theta=0$ case, there is only one type of topological phase. The phase that had two Majorana zero modes becomes trivial upon adding the phase gradient.
The natural question is then what happens to the phases with two Majorana edge states? Despite the fact that the phase has become trivial, one finds that it is still an interesting trivial phase, as was already observed in \onlinecite{bena13}. Here, we study the zero modes of the model, and shed light on the zero mode present in one of the trivial phases.

By transforming $c_j\rightarrow e^{i j \frac{\nabla \theta}{2}} c_j$ as in the previous section, the
Hamiltonian becomes
 \begin{align}
 \label{eq:gen hamil}
H&=\dfrac{t}{2}\sum_{j=1}^{N-1}(e^{i \frac{\nabla \theta }{2}}c^{\dagger}_jc_{j+1}+e^{- i \frac{\nabla \theta }{2} }c^{\dagger}_jc^{\dagger}_{j+1}+h.c.)  \nonumber \\
& + \dfrac{\lambda}{2}\sum_{j=1}^{N-2}(e^{i \nabla \theta}c^{\dagger}_jc_{j+2}+e^{- i \nabla \theta}c^{\dagger}_jc^{\dagger}_{j+2}+h.c.) \nonumber \\
&-\mu \sum_{j=1}^{N}(c^{\dagger}_jc_{j}-\dfrac{1}{2}). 
\end{align}

As we show in the next section (where we consider the model for general parameters), the locations
where the gap of this model closes are very similar to the locations of the phase transitions of the model with zero phase gradient, $\nabla\theta = 0$. Namely, they take the same form, if written in terms of the variables $\tilde{t} = \cos(\nabla\theta/2)t$ and $\tilde{\lambda} = \cos(\nabla\theta)\lambda$, while $\mu$ remains unchanged. So, the gap closes when $\tilde\lambda = \mu \pm \tilde t$, as well as when both $\tilde\lambda = -\mu$ and $|\tilde t| \leq 2 |\mu|$. 

\begin{figure}[t]
\includegraphics[width=.9\columnwidth]{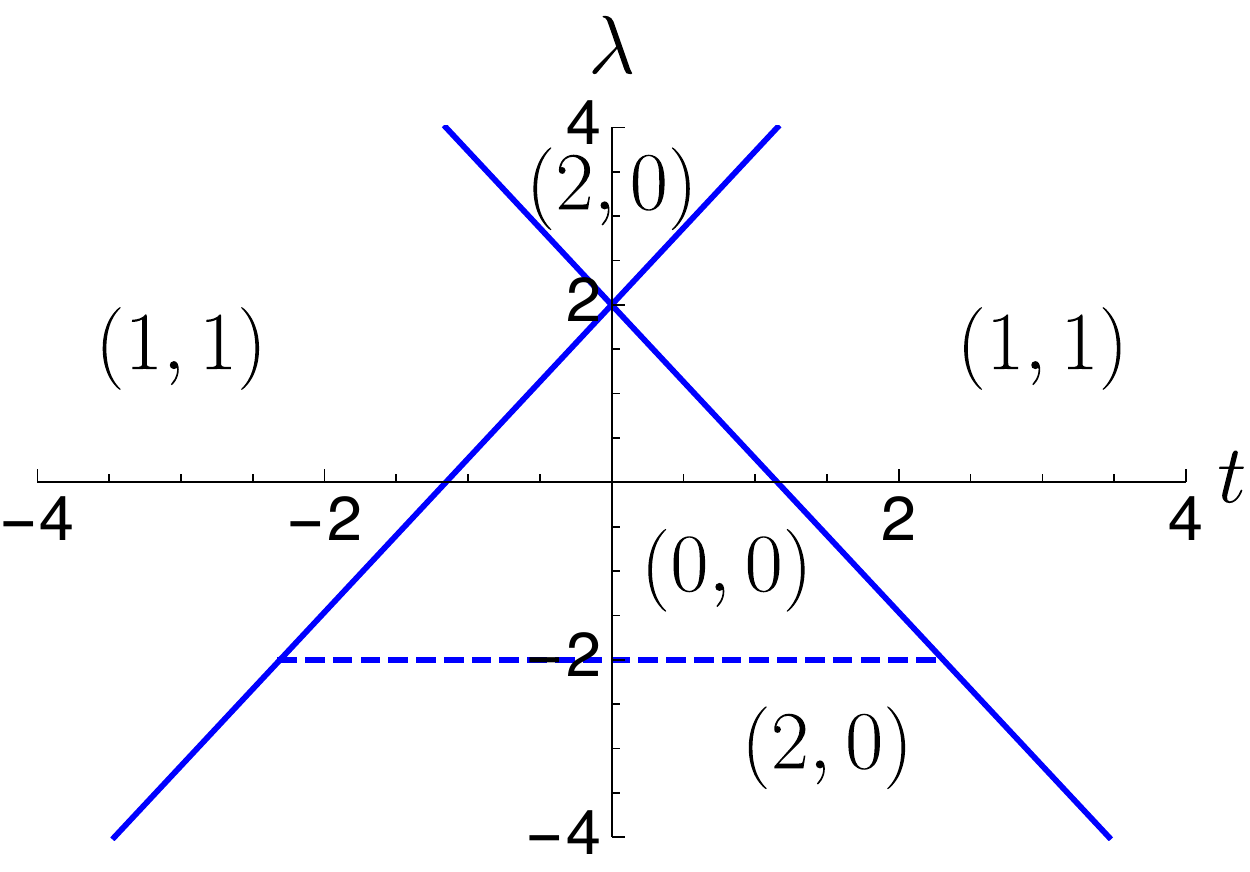}
\caption{Phase diagram for Kitaev chain at $\mu=1$ with NNN terms
and a phase gradient $\nabla\theta = \pi/3$. The numbers in parenthesis in the plot show the number of Majorana zero modes at the left and right side of the system. The solid lines represent the phase boundaries. The dashed line represents a crossover at which the gap closes. The $\lambda$-axis should be treated differently. For $|\lambda| > 2$ and $t=0$ the system is in the topological phase and hosts two Majorana zero modes on each edge, while for $|\lambda| < 2$ and $t=0$, it is in the trivial phase without any zero mode. The points $|\lambda| = 2$ and $t=0$ correspond to phase transitions.}  
\label{fig:phase-diagram-gradient}
\end{figure}

In Fig.~\ref{fig:phase-diagram-gradient}, we show the phase diagram for a phase gradient $\nabla\theta = \pi/3$, and indicate the number of zero-modes on the left and the right separately. We first present the analysis of the zero modes and based on those results we explain the phase diagram of the model.

Sticlet et al.\cite{bena13} showed that the topological phase of this model has one zero mode at both edges as expected. The trivial phase, however, is divided into two regions. One region does not have any zero mode, while the other region has two `Majorana' zero modes that are localized on one edge (i.e., a localized fermionic zero mode), while there is no zero mode on the other edge (see Fig.~\ref{fig:phase-diagram-gradient}).
The former trivial region corresponds to the trivial phase of the model without phase gradient while the later trivial region corresponds to the topological phase of the model without phase gradient with two Majorana zero modes on both sides. In what follows we present analytical wave functions for all the zero modes and determine for which parameters they are present.
To find the Majorana zero modes we use Eqs.~\eqref{eq:LSMreal} and \eqref{eq:LSMimag} and as before, we set $g_n=\dfrac{1}{2}(\phi_n + i \tilde{\psi}_n)$.
The `bulk equations' read
\begin{align}
\label{eq:bulk psi}
&-\mu \tilde{\psi}_n + t\cos(\dfrac{\nabla \theta}{2}) \tilde{\psi}_{n-1}+\lambda \cos(\nabla \theta) \tilde{\psi}_{n-2}=0,\\
\label{eq:bulk phi}
&-\mu \phi_n + t\cos(\dfrac{\nabla \theta}{2}) \phi_{n+1} + \lambda \cos(\nabla \theta) \phi_{n+2}= \nonumber \\
& t \sin(\dfrac{\nabla \theta}{2}) \Big(\tilde{\psi}_{n-1}-\tilde{\psi}_{n+1}\Big)+\lambda \sin(\nabla \theta) \Big(\tilde{\psi}_{n-2}-\tilde{\psi}_{n+2}\Big).
\end{align}
In this case, there are four boundary equations (two for each end) that differ from the bulk ones, namely
\begin{align}
\label{eq:boundary psi}
-\mu \tilde{\psi}_1 &= 0  & t \cos(\dfrac{\nabla \theta}{2}) \tilde{\psi}_1 - \mu \tilde{\psi}_2 &= 0 
\end{align}
and
\begin{align}
\label{eq:boundary phi}
&-\mu \phi_{N-1} + t \cos (\dfrac{\nabla \theta}{2}) \phi_{N} = \\ 
\nonumber &
t \sin (\dfrac{\nabla \theta}{2}) (\tilde{\psi}_{N-2} - \tilde{\psi}_{N})  + \lambda \sin(\nabla \theta) \tilde{\psi}_{N-3} \\
\nonumber
& -\mu \phi_{N} = t \sin (\dfrac{\nabla \theta}{2}) \tilde{\psi}_{N-1} + \lambda \sin(\nabla \theta) \tilde{\psi}_{N-2} \ .
\end{align}
We start by solving the bulk equations, without paying attention to the boundary equations. We then solve the boundary equations, in the different regimes of the phase diagram.

The equations for $\phi_n$ involves the solution for $\psi_n$. Thus, the solution for $\phi_n$
consists of two pieces, namely the general solution to Eq.~\eqref{eq:bulk phi} with the right
hand side set to zero, which we will denote by $\phi_{{\rm gen},n}$,
as well as a specific solution, for the full equation. We start with the ansatz
$\phi_{{\rm gen},n} \sim x^n$. This gives us two correlation lengths
\begin{equation}
\label{eq:cor length scale}
x_{\pm}=\dfrac{-t \cos(\dfrac{\nabla \theta}{2})\pm \sqrt{t^2\cos^2(\dfrac{\nabla \theta}{2})+4\lambda \mu \cos(\nabla \theta)}}{2\lambda \cos(\nabla \theta)} \ .
\end{equation}
Thus, the generic solution is $\phi_{{\rm gen},n} = L_+ x_+^n + L_- x_-^n$, where $L_\pm$ are
constants.
As before, $\tilde{\psi}_{n}=\phi_{{\rm gen},N-n+1}$, which shows that the generic solution for
$\phi$ is localized on the left edge and the solution for $\tilde{\psi}$ is localized on the right edge,
$\tilde{\psi}_n = R_+ x_+^{N-n+1} + R_- x_-^{N-n+1}$, with $R_\pm$ two constants.
To find the full solution $\phi_n$, based on Eq.\eqref{eq:bulk phi}
we need to add a particular solution to $\phi_{{\rm gen},n}$
of the form $S_+ x^{N-n+1}_+ + S_- x^{N-n+1}_- $ with constant
$S_\pm = \kappa_{\pm} R_\pm$, since it should
behave as $\tilde{\psi}$. After some algebra, one finds that
\begin{equation}
\label{eq:kappa}
\kappa_{\pm} =  - \tan(\nabla \theta) + \frac{t \sin(\frac{\nabla \theta}{2})}{\cos(\nabla \theta)\Big[\mu + \lambda \cos(\nabla \theta)\Big]} x_{\pm} \ .
\end{equation}
The general solution to the bulk equations \eqref{eq:bulk psi} and \eqref{eq:bulk phi} is thus given by
\begin{align}
\label{eq:gensolution}
&\tilde{\psi}_n= R_+ x^{N-n+1}_+ + R_- x^{N-n+1}_-,\nonumber \\
&\phi_n= L_+ x^n_+ + L_- x^n_- +  S_+ x^{N-n+1}_+ + S_- x^{N-n+1}_- ,\nonumber \\
& S_{\pm}= \kappa_{\pm} R_{\pm} \ .
\end{align}

With the general solution for the bulk equations at hand, we turn our attention to the boundary
equations, which we solve in the different regimes.
\\
\\
\textit{1) $|x_\pm| >1$}:
In this case, both characteristic length scales are bigger than one, which occurs for the part
of the phase diagram where the model without phase gradient is in the trivial phase. In this case,
it is not hard to convince oneself that the boundary equations \eqref{eq:boundary psi}
and \eqref{eq:boundary psi} lead to $L_\pm = R_\pm = 0$, which means that, as expected, there
are no zero modes in this regime.
\\
\\
\textit{2) $|x_+|<1 $ and $|x_-|>1$}:
In this case, the model is topological and $x_-^n$ increases with $n$,
which means that $x_-^n$ is localized on the right edge instead of the left one.
It is therefore convenient to write this solutions as $\tilde{L}_- (\dfrac{1}{x_-})^{N-n+1}$,
with $\tilde{L}_- = x_-^{N+1} L_-$, to highlight that this solution is localized on the right edge.

The boundary equations \eqref{eq:boundary psi} imply that $R_- = S_- = 0$. 
The boundary equations \eqref{eq:boundary phi} give, after some algebra, that
$\tilde{L}_- = -R_+ \frac{t \sin(\nabla \theta/2)}{\cos(\nabla\theta)(\mu + \lambda \cos(\nabla \theta))}$,
while $S_+ = \kappa_+ R_+$ as before. Thus, the solution for the zero mode is given by
\begin{align}
&\tilde{\psi}_n= R_+ x^{N-n+1}_+ \ , \\
& \phi_n = L_+ x^n_+ + S_+  x^{N-n+1}_+ + \tilde{L}_- (\dfrac{1}{x_-})^{N-n+1} \ .
\end{align}
We find that in this case, there is one zero mode, that is localized on both edges of the system.
One special property of this zero mode, which differs from the case without a phase gradient, is that
$\phi_n$ has support on both edges of the system, while $\psi_n$ only has support on the right edge.
Finally, we note that the case $|x_+| > 1 $ and $|x_-| < 1$ is completely analogous.

\textit{3) $|x_{\pm}|<1$}: This case corresponds to the part of the phase diagram in which the model
without phase gradient has two zero modes on both sides of the system. With the phase gradient,
this model is in a trivial phase. To determine if there are any zero modes, we again solve the
boundary equations.
The boundary equations for $n=1,2$, i.e. \eqref{eq:boundary psi}, give rise to terms that are
proportional to $\tilde{\psi}$ at the left edge. Eq.\eqref{eq:gensolution} assures that these terms
are of order $x_{\pm}^N$ and can be neglected in the thermodynamic limit. So the solution satisfies the boundary equations \eqref{eq:boundary psi}. The boundary equations \eqref{eq:boundary phi} do
give a non-trivial constraint. Namely, for a non-zero phase gradient $\nabla \theta \neq 0$ (for
$\nabla \theta = 0$ the boundary equations are satisfied), one finds that
\begin{align}
R_+ x_+ + R_- x_- &=0 & 
R_+ x^2_+ + R_- x^2_- &=0.
\end{align}
These two boundary equations imply that $R_{\pm}=0$. We conclude that in this regime there are two zero modes on the left side of the system, and none on the right side, i.e. $\phi_n=  L_+ x^n_+ + L_- x^n_-$ and $\tilde{\psi}_n=0$. This precisely corresponds to the surprising result 
obtained by Sticlet et al\cite{bena13}. We stress that this ordinary, or `Dirac' zero mode on the
left side of the system is
not topological, but is in fact a consequence of fine tuning the parameters. We discuss this fine tuning in more detail in Sec.~\ref{sec:general parameters }. Nevertheless, as long as one keeps these parameters fine tuned, the only way this localized zero mode can disappear is via a closing of the gap, signaling a phase transition or a crossover. 


We can now shed light on the phase diagram Fig.~\ref{fig:phase-diagram-gradient}
(where we set $\nabla\theta = \pi/3$) and discuss it in more detail. The two solid lines in the figure,
$\lambda = \mu / \cos(\nabla \theta)  \pm t \cos (\nabla \theta/2)/\cos(\nabla \theta)$,
indicate phase transitions. Along the dashed line, $\lambda =- \mu/ \cos (\nabla \theta)$, the gap closes and a crossover occurs between a region with no zero mode and a region where the system has two zero modes on left edge and none on the right edge.
The corners of the triangular region without any zero modes are given by
$(0,\mu/ \cos (\nabla \theta))$ and $(\pm 2\mu/ \cos (\nabla \theta/2),-\mu/ \cos (\nabla \theta))$.

Upon increasing $\nabla \theta$, the slope of the lines indicating the phase transitions increases, and
the size of the trivial center region increases.
At $\nabla \theta = \pi/2$, the two phase transition lines are parametrized by $t= \pm \sqrt{2} \mu$ and the trivial center region without any zero mode becomes a stripe in the middle of phase diagram. For $\nabla \theta \in (\pi/2,\pi)$ the shape of the phase diagram is inverted with respect to the $t$-axis in comparison with Fig.~\ref{fig:phase-diagram-gradient}.

We should note that the $\lambda$-axis ($t=0$) should be treated separately. For $t=0$, the model corresponds to two copies of the model that we studied in Sec.~\ref{sec Pfeutyphase}, one for the even sites and another one for the odd sites. Therefore, the system is either in a trivial phase (for $t=0$ and $| \mu | > | \lambda \cos(\nabla \theta)|$) or in a topological phase (for $t=0$ and $| \mu | < | \lambda \cos(\nabla \theta)|$). In the topological phase the system has two Majorana zero modes on each edge, because the two chains are decoupled. 

\begin{figure}[t]
\includegraphics[width=.9\columnwidth]{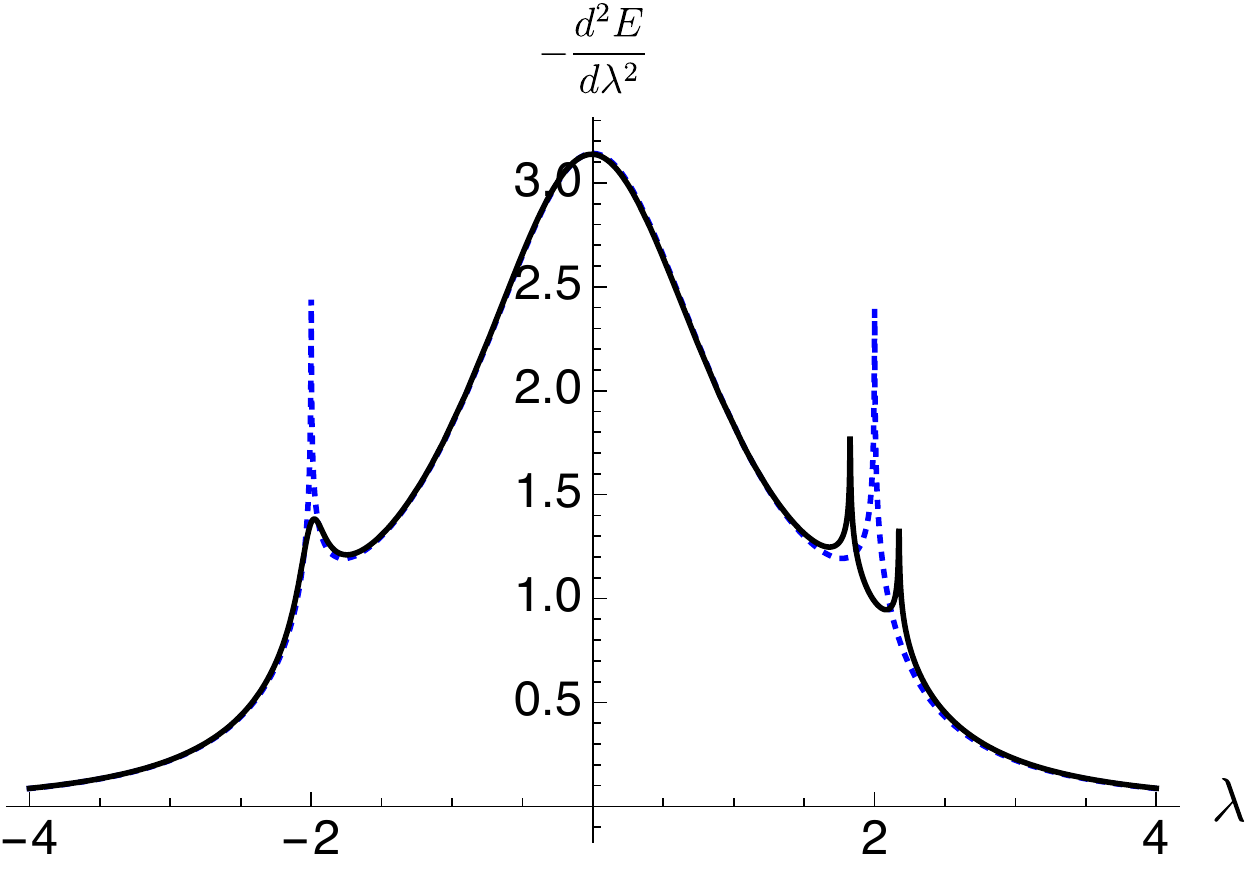}
\caption{The second derivative of the ground state energy as a function of $\lambda$ for $\mu =1$ and $\nabla \theta = \pi/3$. We set $t=0$ for the dashed line and $t=0.1$ for the solid line.}  
\label{fig:2nd derivative}
\end{figure} 

To provide further insight in the phase diagram (Fig.~\ref{fig:phase-diagram-gradient}), we calculated the second derivative of the ground state energy for $\mu=1$, $\nabla \theta = \pi/3$ and $t=0$ as well as $t=0.1$ for $\lambda \in [-4,4]$\footnote{We used the version of the model with periodic boundary conditions, Eq.~\ref{eq:kspace hamil}, that will be presented in Sec.\ref{sec:general parameters }.}. The result is shown in Fig.~\ref{fig:2nd derivative}. We first consider the line $t=0$ (the blue dashed line). As we described above, for $\lambda=-4$ the system is in the topological phase. The point $\lambda=-2$ is a critical point. Upon increasing $\lambda$, one enters the trivial phase. By passing the other critical point, namely $\lambda=+2$, one enters another topological phase. These two critical points give rise to divergencies in $-d^2E/d\lambda^2$ as is shown in Fig.~\ref{fig:2nd derivative}. This is a clear signature of a second order phase transition. For $t\neq 0$, (in the figure we used $t=0.1$, the black line), the situation is quite different. At $\lambda=-2$, $-d^2E/d\lambda^2$ is smooth (the critical point at $\lambda = -2$ and $t=0$ shows its presence via a bump) and the system undergoes a crossover, although the gap closes. One still observes two divergencies symmetrically around $\lambda = 2$.
These two divergencies correspond to the phase transitions indicated by the solid lines in Fig.~\ref{fig:phase-diagram-gradient}.

We close this section by mentioning that it is of course possible to have the localized Dirac zero mode
at the other edge of the system. One way to achieve this is by changing the phase dependence of the original pairing terms in the original Hamiltonian Eq.~\eqref{eq:hamil phase gradient with NNN} to
$\frac{t}{2} e^{ i(j+1) \nabla \theta }c^{\dagger}_jc^{\dagger}_{j+1}+ \frac{\lambda}{2} e^{ i(j+2) \nabla \theta }c^{\dagger}_jc^{\dagger}_{j+2}$. This is equivalent to an inversion accompanied by the change $\nabla\theta \rightarrow -\nabla\theta$. We note that merely changing
$\nabla\theta \rightarrow -\nabla\theta$ does not change the role of the left and right hand side of the
system. Basically the same calculation as above shows that this new pairing leads to two zero modes on right edge and none on the left edge (because the role of $\phi_n$ and $\psi_n$ in Eqs.~\eqref{eq:bulk psi} and \eqref{eq:bulk phi} is swapped). The model with these pairing terms has the same topological and trivial phases, however the left and right sides of the chain change their role.  As we show in the next subsection, in the translational invariant formulation of the model, as in Eq.~\eqref{eq:gen hamil}, the location of the zero-modes is determined by the relative sign of the phases of the hopping and paring terms.

\subsection{The general case}
\label{sec:general parameters }
To understand the fine tuning that is necessary to have the Dirac zero mode that resides
on one side of the system as described in the previous section, 
we look at the more general Hamiltonian,
  \begin{align}
 \label{eq:General NNN}
H&=\dfrac{1}{2}\sum_{j=1}^{N-1}(t_1c^{\dagger}_jc_{j+1}+\Delta_1 c^{\dagger}_jc^{\dagger}_{j+1}+h.c.)  \nonumber \\
& + \dfrac{1}{2}\sum_{j=1}^{N-2}(t_2 c^{\dagger}_jc_{j+2}+\Delta_2 c^{\dagger}_jc^{\dagger}_{j+2}+h.c.) \nonumber \\
&-\mu \sum_{j=1}^{N}(c^{\dagger}_jc_{j}-\dfrac{1}{2}),
\end{align}
where $t_1$, $\Delta_1$, $t_2$ and $\Delta_2$ are arbitrary complex parameters. 
In the case of periodic boundary conditions, we can define
$\Psi_k=(c_k,c^{\dagger}_{-k})^\mathrm{T}$ and write the Hamiltonian as:
\begin{align}
\label{eq:kspace hamil}
H&= \dfrac{1}{2}\sum_k \Psi^{\dagger}_k \mathcal{H}_k \Psi_k,\ \mathcal{H}_k=h_0(k)\mathbf{1}+\mathbf{h}(k).\bm{\tau}, \nonumber \\
h_0(k)&= -\Im(t_1) \sin k - \Im(t_2) \sin(2k), \nonumber \\
h_1(k) &= -\Im(\Delta_1) \sin k - \Im(\Delta_2) \sin(2k), \nonumber \\
h_2(k)&= -\Re(\Delta_1) \sin k - \Re(\Delta_2) \sin(2k), \nonumber \\
 h_3(k)&=\Re(t_1) \cos k + \Re(t_2) \cos(2k)-\mu,
\end{align}
where $\mathbf{1}$ is the two by two identity matrix.
Performing a unitary transformation with $U= \frac{1}{\sqrt{2}}(\tau^x + \tau^z)$ we get,
\begin{equation}
\label{eq:Qmatrix}
\mathcal{Q}_k = U^{\dagger} \mathcal{H}_k U=
\left(
\begin{array}{cc}
h_0(k)+h_1(k) &h_3(k) + i h_2(k) \\
h_3(k) - i h_2(k) &  h_0(k) - h_1(k)
\end{array} \right).
\end{equation}
By comparing the model we discuss here, Eq.~\eqref{eq:gen hamil} with \eqref{eq:kspace hamil}
we find that in this case, all the imaginary parts depends are proportional to $\sin(\nabla \theta/2)$
or $\sin(\nabla \theta)$, for the nearest neighbor or NNN case respectively. So, these terms
vanish for $\nabla \theta=0$. In that case, we obtain
\begin{equation}
\mathcal{Q}_k\big|_{\nabla \theta=0} = 
\left(
\begin{array}{cc}
0 &h_3(k) + i h_2(k) \\
h_3(k) - i h_2(k) &  0
\end{array} \right).
\end{equation}
Since we performed an unitary transformation, $\mathrm{Det} \ \mathcal{Q}_k= \mathrm{Det} \ \mathcal{H}_k$. So in the gapped phase, either topological or trivial, $\mathrm{Det} \mathcal{Q}_k =|h_3(k) + i h_2(k)|^2 \neq 0$ can be used to define a topological invariant via the winding of
$\mathrm{Arg} (h_3(k) + i h_2(k))$, see \cite{ryu10}.
This calculation leads to the same phase diagram we discussed before, see Fig.~\ref{fig: PDNNN}. 

We now consider a phase gradient, i.e. $\nabla\theta \neq 0$,
which is the case we are interested in. Based on Eq.~\eqref{eq:gen hamil} we have
\begin{align}
h_0(k)&= -t \sin(\frac{\nabla \theta}{2}) \sin k - \lambda \sin(\nabla \theta) \sin(2k), \nonumber \\
h_1(k) &= t \sin(\frac{\nabla \theta}{2}) \sin k +\lambda \sin(\nabla \theta) \sin(2k),\nonumber \\
h_2(k)&= -t \cos(\frac{\nabla \theta}{2}) \sin k - \lambda \cos(\nabla \theta) \sin(2k), \nonumber \\
 h_3(k)&=t \cos(\frac{\nabla \theta}{2}) \cos k + \lambda \cos(\nabla \theta) \cos(2k)-\mu.
\end{align}
The fact that $\Im(t_1)+\Im(\Delta_1)= \Im(t_2)+\Im(\Delta_2)=0$, gives rise to $\mathcal{Q}_{k,11}=0$. This means that, similar to the $\nabla \theta=0$ case, we have that
$\mathrm{Det} \mathcal{Q}_k =|h_3(k) + i h_2(k)|^2$, despite the fact that $\mathcal{Q}_{k,22}\neq 0$.

Thus we find \textit{effectively} the same phase diagram for the model with the phase gradient,
namely the one given in Fig.~\ref{fig: PDNNN}, if we replace
$t \rightarrow t \cos(\frac{\nabla \theta}{2})$ and $\lambda \rightarrow \lambda \cos(\nabla \theta)$.

As we indicated in the previous section, by changing the paring terms in the original Hamiltonian
Eq.~\eqref{eq:hamil phase gradient with NNN} to 
$
\frac{t}{2} e^{ i(j+1) \nabla \theta } c^{\dagger} _jc^{\dagger}_{j+1}
+
\frac{\lambda}{2} e^{ i(j+2) \nabla \theta }c^{\dagger}_jc^{\dagger}_{j+2}
$,
we can have the situation that the system has two `Majorana' zero modes on the right edge and
none on the left side.
The gauge transformation
$\tilde{c}_j=e^{i j \frac{\nabla \theta}{2}} c_j$ changes these terms to
$\frac{t}{2} e^{ i \frac{\nabla \theta }{2} }\tilde{c}^{\dagger}_j \tilde{c}^{\dagger}_{j+1}+ \frac{\lambda}{2} e^{ i \nabla \theta  }\tilde{c}^{\dagger}_j\tilde{c}^{\dagger}_{j+2}$.
The hopping terms $\frac{t}{2}c^\dagger_j c_{j+1} + \frac{\lambda}{2} c^\dagger_j c_{j+2}$
become
$
\frac{t}{2} e^{ i \frac{\nabla \theta }{2} }\tilde{c}^{\dagger}_j \tilde{c}_{j+1}
+
\frac{\lambda}{2} e^{ i \nabla \theta} \tilde{c}^{\dagger}_j\tilde{c}_{j+2}$
as before.
In this case we find that $\Im(t_1)-\Im(\Delta_1)= \Im(t_2)-\Im(\Delta_2)=0$,
which results in $\mathcal{Q}_{k,22}=0$.

In class BDI, all the information about the zero modes is encoded in $\mathrm{Det} \mathcal{Q}_k$.
The discussion above shows that this is not so in the present case.
Whether $\mathcal{Q}_{k,11}=0$ or $\mathcal{Q}_{k,22}=0$ plays an important role in
determining the position of the (non-topological) localized zero modes.
It is also clear what fine tuning we need in order to have a pair of `Majorana' zero modes
localized at one side of the system and none on the other.
We need either $\mathcal{Q}_{k,11}$ or $\mathcal{Q}_{k,22}$ to be zero, but not both.
This is the case if we fine tune
$\Im(t_1+\Delta_1)=\Im(t_2+\Delta_2)=0$ or $\Im(t_1- \Delta_1)=\Im(t_2- \Delta_2)=0$, but not
both which would imply that all these parameters are real, and one has an equal number of
zero-modes on either side of the system.
 
To explore this situation further, we write the Hamiltonian in terms of
Majorana operators,
\begin{align}
\label{eq:Majorana hamil}
H&=  \frac{\Im(t_1+\Delta_1)}{4}\sum_{j=1}^{N-1} i \gamma_{A,j}\gamma_{A,j+1} \nonumber \\
& + \frac{\Re(t_1-\Delta_1)}{4}\sum_{j=1}^{N-1} i \gamma_{A,j}\gamma_{B,j+1} \nonumber \\
& - \frac{\Re(t_1+\Delta_1)}{4}\sum_{j=1}^{N-1} i \gamma_{B,j}\gamma_{A,j+1} \nonumber \\
& + \frac{\Im(t_1-\Delta_1)}{4}\sum_{j=1}^{N-1} i \gamma_{B,j}\gamma_{B,j+1} \nonumber \\
 & + \frac{\Im(t_2+\Delta_2)}{4}\sum_{j=1}^{N-2} i \gamma_{A,j}\gamma_{A,j+2} \nonumber \\
& + \frac{\Re(t_2-\Delta_2)}{4}\sum_{j=1}^{N-2} i \gamma_{A,j}\gamma_{B,j+2} \nonumber \\
& - \frac{\Re(t_2+\Delta_2)}{4}\sum_{j=1}^{N-2} i \gamma_{B,j}\gamma_{A,j+2} \nonumber \\
& + \frac{\Im(t_2-\Delta_2)}{4}\sum_{j=1}^{N-2} i \gamma_{B,j}\gamma_{B,j+2} \nonumber \\
&-\frac{\mu}{2}\sum_{j=1}^N i \gamma_{A,j}\gamma_{B,j}.
\end{align}

As a simple example we can see that for the Hamiltonian presented in Eq.~\eqref{eq:gen hamil}, setting $\mu =0$ and $\nabla \theta=\pi$, yields following Majorana representation
\begin{equation}
H =  \frac{t}{2}\sum_{j=1}^{N-1} i \gamma_{B,j}\gamma_{B,j+1}+ \frac{\lambda}{2}\sum_{j=1}^{N-2} i \gamma_{B,j}\gamma_{A,j+2},
\end{equation}
where it is evident that $\gamma_{A,1}$ and $\gamma_{A,2}$ do not appear in the Hamiltonian and therefore commute with it. Hence there are two Majorana zero modes on the left edge.

It is interesting to note that for $\nabla \theta = \pi$, the Hamiltonian belongs to class
BDI. From the form of the Hamiltonian in Eq.~\eqref{eq:gen hamil} this is not obvious,
but it is for the form Eq.~\eqref{eq:hamil phase gradient with NNN}, because all the coupling
constants are real. On the other hand, in this form, some of the couplings are staggered,
and in the periodic case, the model is translationally invariant with a two-site unit cell. The
phase diagram has a different structure in this case, with only three phases. The phase
boundaries do not depend on $t$, and are given by $\lambda = \pm \mu$. Because of the
two-site unit cell, we use the formulation of the phase-winding invariant as given by \cite{tewari12}.
One finds that all three phases are in fact trivial. In the trivial phases with $|\lambda| > |\mu|$, there
is a localized Dirac zero mode only on the left side of the system, and no zero modes on the right side.
This is consistent with the analysis of the model based on Eq.~\eqref{eq:gen hamil}.
From the point $\nabla \theta = \pi$ it is clear that
also in symmetry class BDI, there are Hamiltonians that have trivial phases, which have a localized
Dirac zero mode only on one side of the system, if parameters are fine-tuned.

Our previous discussion led us to conclude that $\Im(t_1 +\Delta_1)=\Im(t_2+ \Delta_2)=0$ could result in two zero modes on the left edge. Based on Eq.~\eqref{eq:Majorana hamil} we can see that this means that there should not be any terms like $i\gamma_{A,j}\gamma_{A,j+1}$ and
$i\gamma_{A,j}\gamma_{A,j+2}$ present in the Hamiltonian.
We can shed more light on this issue based on our analytical solution for the  non-uniform pairing with nearest neighbor hopping and pairing. 

As a first a step, we assume that $t=0$. This means that we have two decoupled chains with a phase gradient. Our previous analysis shows that in the topological phase we have one Majorana zero mode on each edge. The wavefunctions for these Majorana modes are given in
Eqs.~\eqref{eq:Pfeuty + phi left} and \eqref{eq:Pfeuty + phi right}.
The crucial difference between these two wave functions originates in the direction of the phase gradient, which causes the left mode to be independent of $\gamma_B$, while the right modes consists of both $\gamma_A$ and $\gamma_B$.
Namely, for the left mode $g_n$ is purely real while for the right mode $g_n$ is complex, and hence involves both $\phi$ and $\tilde{\psi}$. 

In the second step, we turn on nearest neighbor couplings, i.e. $t \neq 0$. We see that the first four terms in the Hamiltonian Eq.~\eqref{eq:Majorana hamil} result in a coupling between the zero modes
of the two decoupled chains. Under the assumption that $\Im(t_1 + \Delta_1)=0$ and
$\Im(t_2 + \Delta_2)=0$, which holds in our analytic calculation of the zero modes, we find that
the right zero modes from the two different chains become coupled because of the
$i \gamma_{B,j} \gamma_{B,j+1}$ terms present then $t\neq 0$, which gaps them out. 
On the other hand, the zero modes on the left edge do not become coupled directly, and remain
gapless. Their wavefunctions are modified to the new ones presented in Eq.~\eqref{eq:gensolution}.

Finally, we mention that we checked numerically that under the conditions
$\Im(t_1 + \Delta_1)=\Im(t_2+ \Delta_2)=0$, the system has a phase for which the
ground state has two zero modes located on the left edge, and none on the right edge.
The same holds true in the case that $\Im(t_1 - \Delta_1)=\Im(t_2- \Delta_2)=0$, if one
exchanges the left and right edge of the system.

\section{Discussion}
\label{sec:discussion}

In this paper, we investigated the `one-sided' fermionic zero modes observed by
Sticlet et al.\cite{bena13}, by solving the Kitaev model, in the presence of complex hopping and pairing terms, including NNN terms, for open chains. From our investigation, it became clear that fine-tuned parameters are necessary for such zero modes to exist, but under the fine-tuned conditions, the gap
needs to close in order to destroy them. Leaving the fine-tuned conditions gaps these zero-modes
out, turning them into one-sided low-energy subgap modes. Phases with such one-sided bound states
can occur both in one-dimensional systems in class D, as well as in class BDI.
These modes are not protected by topology, which means that they can
occur in the topologically trivial phase.

The general condition for the existence of `one-sided zero modes' is most easily explained
in terms of the Majorana formulation of the chains. Starting from a situation in which two pairs of
(delocalized) Majorana bound states are present (i.e., in class BDI), one needs a perturbing term
such that the two Majoranas describing the mode on, say, the left side are coupled, while the modes
on the right side are not. It is worth to mention that if one assumes a different phase gradient for the nearest neighbour and next-nearest neighbour pairing terms in Eq.~\ref{eq:hamil phase gradient with NNN}, these `one-sided zero modes' gap out.

There has been a lot of progress on models in higher dimensions, that exhibit exact zero modes, see
for instance Ref.~\onlinecite{kunst17}.
It would be interesting to investigate if it is possible to construct models, that exhibit `one-sided' zero
modes along the lines of the ones described in this paper, even in those higher-dimensional systems.
\\
\\
{\em Acknowledgements --}
We would like to thank F.~Pollmann, C. Sp\r{a}nsl\"att and R. Verresen for
interesting discussions. This research was sponsored, in part, by the Swedish research council.

\appendix

\section{Details of the Kitaev chain spectrum calculation }
\label{sec:Appendix GKC}
In this appendix we present details of the solution for the open Kitaev chain with generic real
parameters and free boundary conditions. The Hamiltonian reads:
\begin{align}
\label{eq:appA kitaev hamil}
H&=\dfrac{1}{2}\sum_{j=1}^{N-1}(c^{\dagger}_jc_{j+1}+h.c) +\dfrac{\Delta}{2}\sum_{j=1}^{N-1}(c^{\dagger}_jc^{\dagger}_{j+1}+h.c)\nonumber \\
&-\mu \sum_{j=1}^{N}(c^{\dagger}_jc_{j}-\dfrac{1}{2}).
\end{align}
It is helpful to recall the (of course well known) solution for the periodic case \cite{}, which is obtained 
via a Fourier transformation,
$c_j=\dfrac{1}{\sqrt{N}} \sum_k e^{i k j} c_k$, and defining $\Psi_k=(c_k,c^{\dagger}_{-k})^\mathrm{T}$. This results in 
\begin{equation}
H = \dfrac{1}{2}\sum_k \Psi_k^{\dagger} 
\left(
\begin{array}{cc}
-\mu+\cos k & i \Delta \sin k \\
- i \Delta \sin k &  \mu- \cos k
\end{array} \right)
\Psi_k.
\end{equation}
Diagonalization of this $2\times 2$ matrix gives us:
\begin{eqnarray}
H=\sum_k \epsilon_k(f^{\dagger}_k f_k - \dfrac{1}{2}), \\
\label{eq:appA dispersion}
\epsilon_k=\sqrt{(\mu - \cos k)^2+\Delta^2 \sin^2k},
\end{eqnarray}
 where $f_k$ is a new fermionic quasiparticle annihilation operator.
 
To tackle the open case, 
we use the LSM method which is reviewed in Sec.~\ref{secLSM}.
To this end, we need to arrange the Hamiltonian to have the form of Eq.~\eqref{eq:LSM convention},
\begin{equation}
H= \sum_{i,j=1}^N c^{\dagger}_i A_{ij}c_j +\dfrac{1}{2}(c^{\dagger}_i B_{ij}c^{\dagger}_j +h.c. ).
\end{equation}  
To find $\phi_{\alpha}$ and $\psi_{\alpha}$ from Eqs.~\eqref{eq:finding phi} and \eqref{eq:finding psi}
i.e.,
\begin{eqnarray*}
\phi_{\alpha} (A-B)(A+B) &=& \Lambda_{\alpha}^2 \phi_{\alpha} \\
\psi_{\alpha} (A+B)(A-B) &=& \Lambda_{\alpha}^2 \psi_{\alpha} \ ,
\end{eqnarray*}
we have to construct the matrices $A-B$ and $A+B$.
 
We present these matrices for the more general case of Hamiltonian in Eq.~\eqref{eq:General NNN},
i.e.,
\begin{align}
 H&=\dfrac{1}{2}\sum_{j=1}^{N-1}(t_1c^{\dagger}_jc_{j+1}+\Delta_1 c^{\dagger}_jc^{\dagger}_{j+1}+h.c.)  \nonumber \\
& + \dfrac{1}{2}\sum_{j=1}^{N-2}(t_2 c^{\dagger}_jc_{j+2}+\Delta_2 c^{\dagger}_jc^{\dagger}_{j+2}+h.c.) \nonumber \\
&-\mu \sum_{j=1}^{N}(c^{\dagger}_jc_{j}-\dfrac{1}{2}),
\end{align}
because we need them later on. In this case,
$A-B$ and $A+B$ read,
\begin{widetext}
\begin{eqnarray}
\label{eq:appA gen A-B A+B}
A-B =\dfrac{1}{2} 
\begin{pmatrix}
-2\mu & t_1-\Delta_1 &t_2 - \Delta_2 & & & &  & &   \\
t_1^*+\Delta_1 &-2 \mu &t_1-\Delta_1 &t_2 - \Delta_2 &  & &  &\mathbf{0} &   \\
t_2^*+\Delta_2 & t_1^*+\Delta_1 &-2\mu& t_1-\Delta_1 &t_2 -\Delta_2  & &  & &   \\
 & & & &\ddots  & &  & &\\
& & & & t_2^*+\Delta_2& t_1^*+\Delta_1& -2\mu & t_1-\Delta_1&t_2-\Delta_2\\ 
 & \mathbf{0}& & & & t_2^*+\Delta_2& t_1^*+\Delta_1&-2\mu &t_1-\Delta_1\\
 & & & &  & &  t_2^*+\Delta_2 & t_1^*+\Delta_1&-2\mu 
\end{pmatrix}, \\
A+B =\dfrac{1}{2} 
\begin{pmatrix}
-2\mu & t_1+\Delta_1 &t_2 + \Delta_2 & & & &  & &   \\
t_1^*-\Delta_1 &-2 \mu &t_1+\Delta_1 &t_2+ \Delta_2 &  & &  &\mathbf{0} &   \\
t_2^*-\Delta_2 & t_1^*-\Delta_1 &-2\mu& t_1+\Delta_1 &t_2 +\Delta_2  & &  & &   \\
 & & & &\ddots  & &  & &\\
& & & & t_2^*-\Delta_2& t_1^*-\Delta_1& -2\mu & t_1+\Delta_1&t_2+\Delta_2\\ 
 & \mathbf{0}& & & & t_2^*-\Delta_2& t_1^*-\Delta_1&-2\mu &t_1+\Delta_1\\
 & & & &  & &  t_2^*-\Delta_2 & t_1^*-\Delta_1&-2\mu 
\end{pmatrix}.
\end{eqnarray}
For the Hamiltonian Eq.~\eqref{eq:appA kitaev hamil}, i.e. with $t_1=1$, $\Delta_1=\Delta$,
and $t_2=\Delta_2=0$, these reduce to,
\begin{eqnarray}
A-B =\dfrac{1}{2} 
\begin{pmatrix}
-2\mu & 1-\Delta & & & & &  & &   \\
1+\Delta &-2 \mu &1-\Delta& &  & &  &\mathbf{0} &   \\
0& 1+\Delta &-2\mu&1-\Delta&  & &  & &   \\
 & & & &\ddots  & &  & &\\
& & & & & 1+\Delta& -2\mu & 1-\Delta&0\\ 
 & \mathbf{0}& & & & & 1+\Delta &-2\mu &1-\Delta\\
 & & & &  & &  & 1+\Delta&-2\mu 
\end{pmatrix}, \\
A+B =\dfrac{1}{2} 
\begin{pmatrix}
-2\mu & 1+\Delta & & & & &  & &   \\
1-\Delta &-2 \mu &1+\Delta& &  & &  &\mathbf{0} &   \\
0& 1-\Delta &-2\mu&1+\Delta&  & &  & &   \\
 & & & &\ddots  & &  & &\\
& & & & & 1-\Delta& -2\mu & 1+\Delta&0\\ 
 & \mathbf{0}& & & & & 1-\Delta &-2\mu &1+\Delta\\
 & & & &  & &  & 1-\Delta&-2\mu 
\end{pmatrix}.
\end{eqnarray}
Using these matrices in Eq.~\eqref{eq:finding phi} one gets
\begin{equation}
\label{eq: appA bulk equation}
(1-\Delta^2)\phi_{\alpha,n-2} - 4\mu \phi_{\alpha,n-1} + [4\mu^2 +2(1+\Delta^2)]\phi_{\alpha,n}
-4\mu \phi_{\alpha,n+1}+(1-\Delta^2)\phi_{\alpha,n+2} 
=4\Lambda^2_{\alpha}\phi_{\alpha,n}, 
\end{equation}
for $3\leq n\leq N-2$.
We call this the `bulk equation'. In the case of periodic boundary conditions, this is actually the only equation one has to consider.
However, for an open chain with free boundary conditions, we also have four boundary equations which are different from the bulk one, namely for
$n=1,2,N-1$ and $N$ one has:
\begin{align}
\label{eq:appA boundary1}
[4\mu^2 + (1-\Delta)^2]\phi_{\alpha,1} -4 \mu \phi_{\alpha,2}+(1-\Delta^2)\phi_{\alpha,3}&=4\Lambda^2_{\alpha}\phi_{\alpha,1} & (n=1) \\
\label{eq:appA boundary2}
-4\mu\phi_{\alpha,1}+  [4\mu^2 + 2(1+\Delta^2)]\phi_{\alpha,2} -4 \mu \phi_{\alpha,3}+(1-\Delta^2)\phi_{\alpha,4} &=4\Lambda^2_{\alpha}\phi_{\alpha,2} & (n=2) \\
\label{eq:appA boundaryN-1}
(1-\Delta^2)\phi_{\alpha,N-3}-4 \mu \phi_{\alpha,N-2}+  [4\mu^2 + 2(1+\Delta^2)]\phi_{\alpha,N-1}-4\mu\phi_{\alpha,N}&=4\Lambda^2_{\alpha}\phi_{\alpha,N-1} &(n=N-1)\\
\label{eq:appA boundaryN}
(1-\Delta^2)\phi_{\alpha,N-2} -4 \mu \phi_{\alpha,N-1}+[4\mu^2 + (1+\Delta)^2]\phi_{\alpha,N} &=4\Lambda^2_{\alpha}\phi_{\alpha,N} & (n=N) .
\end{align}
\end{widetext}
We note the difference between the $\phi_{\alpha,1}$ term in the equation for $n=1$ and the
$\phi_{\alpha,N}$ term in the equation for $n=N$.

To solve these equations we can start with an ansatz for the eigenvalues $\Lambda_\alpha$. Note that the bulk equation is the same for both the periodic and the open chain. This suggests to use our knowledge about the periodic case.
The bulk equation determines the form of the eigenvalues as a function of a parameter $\alpha$, which in turn is determined by the boundary equations.
This is exactly what happens in the periodic case, where we use $k$ instead of $\alpha$ and fix $k=\dfrac{2\pi n}{N}$ for $n = 0,1,\ldots N-1$ in Eq.~\eqref{eq:appA dispersion} by demanding $c_{N+1}=c_{1}$.
 
Therefore we use the same parametrization for the eigenvalues as in the open case,
 \begin{equation}
 \label{eq: appA energy level}
 \Lambda_{\alpha}^2=(\mu - \cos \alpha)^2+\Delta^2 \sin^2\alpha.
 \end{equation}
Now we need to find an equation based on which one can determine all the possible values of $\alpha$. 
With the ansatz for $\Lambda_{\alpha}$, we solve Eq.~\eqref{eq: appA bulk equation} by
the standard approach, i.e. we consider $\phi_{\alpha,n} \sim x^n_{\alpha}$. 
Using this in Eq.~\eqref{eq: appA bulk equation} gives us:
\begin{align}
&x^4_{\alpha} - K x^3_{\alpha}+2(K \cos \alpha - \cos2\alpha)x^2_{\alpha} - K x_{\alpha} +1=0, \\
&K= \dfrac{4\mu}{1- \Delta^2}.
\end{align}
One checks that $e^{\pm i \alpha}$ are solutions independent of the parameter $K$. Since we have found two roots, we can find the other two, which are given by $e^{\pm i \beta}$ where $\beta$ satisfies
\begin{equation}
\label{eq:appA alpha beta1}
\cos \alpha + \cos \beta =\dfrac{K}{2}.
\end{equation}
Therefore each $\alpha$ has a $\beta$ partner. We note that $\alpha$ and $\beta$ are equivalent. The associated eigenvalues can be written in the same functional form, i.e. $\Lambda_{\alpha}=\Lambda_{\beta}=\sqrt{(\mu - \cos \beta)^2+\Delta^2 \sin^2\beta}$, which
follows from Eq.~\eqref{eq:appA alpha beta1}. We continue to use $\alpha$ as the label indicating the eigenvalue. 

These solutions tell that $e^{\pm i n \alpha}$ and $e^{\pm i n \beta}$ are the most general solution for the bulk equation. Now we need to determine a linear combination of these functions that satisfies the boundary equations. Treating the left and right edges in an equivalent way, we consider the following combination:
\begin{align}
\phi_{\alpha,n}&=A_1 \sin(n\alpha) +A_2 \sin[(N+1-n)\alpha] \nonumber \\
& + B_1 \sin(n\beta) +B_2 \sin[(N+1-n)\beta] \ ,
\end{align}
in which $A_1$,$A_2$,$B_1$ and $B_2$ are constants.

Using this ansatz, Eqs.~\eqref{eq:appA boundary2} and \eqref{eq:appA boundaryN-1} give us:
\begin{align}
&A_1 \sin[(N+1)\alpha] + B_1 \sin [(N+1)\beta]=0  \\
&A_2 \sin[(N+1)\alpha] + B_2 \sin [(N+1)\beta]=0 \ .
\end{align}
Based on these relations, we rewrite the ansatz:
\begin{align}
\label{eq: appA finalansatz}
\phi_{\alpha,n}&=A_1 \Big\{ \sin(n\alpha)
- \frac{\sin[(N+1)\alpha]}{\sin[(N+1)\beta]} \sin(n \beta) \Big\}\nonumber \\
& + A_2 \Big\{  \sin[(N+1-n)\alpha] \nonumber \\
&- \frac{\sin[(N+1)\alpha]}{\sin[(N+1)\beta]} \sin[(N+1-n) \beta] \Big\} .
\end{align}

Finally, we make sure that the ansatz satisfies 
Eqs.~\eqref{eq:appA boundary1} and \eqref{eq:appA boundaryN}, which leads to the following equations:
\begin{widetext}
\begin{eqnarray}
\label{eq:appA A1 A2}
\left(
\begin{array}{cc}
-\dfrac{\Delta}{1-\Delta} f_3(\alpha,\beta) &-\dfrac{\Delta}{1-\Delta} f_1(\alpha,\beta)+f_2(\alpha,\beta) \\
\dfrac{\Delta}{1+\Delta} f_1(\alpha,\beta)+f_2(\alpha,\beta) & \dfrac{\Delta}{1+\Delta} f_3(\alpha,\beta)
\end{array} \right)
 \left(
\begin{array}{c}
A_1 \\
A_2
\end{array} \right)=
 \left(
\begin{array}{c}
0 \\
0
\end{array} \right)
\end{eqnarray}
\end{widetext}
in terms of the functions
\begin{align}
 &f_1(\alpha,\beta)= \sin(N\alpha) - \dfrac{\sin[(N+1)\alpha]}{\sin[(N+1)\beta]} \sin(N\beta) \\
&f_2(\alpha,\beta) = \sin[(N+1)\alpha](\cos \beta - \cos \alpha)\\
&f_3(\alpha,\beta)=\sin\alpha - \dfrac{\sin[(N+1)\alpha]}{\sin[(N+1)\beta]} \sin\beta \ .
\end{align}
 
To find a non-trivial solution for $A_1$ and $A_2$, we require that the determinant of the
matrix in Eq.~\eqref{eq:appA A1 A2} is zero. This gives us another equation for $\alpha$ and $\beta$:
 \begin{align}
 \label{eq:appA alpha beta2}
 &\sin^2\alpha +\sin^2\beta + \dfrac{1}{\Delta^2}(\cos\beta-\cos\alpha)^2\nonumber \\
&-2\dfrac{\sin \alpha \sin \beta}{\sin[(N+1)\alpha]\sin[(N+1)\beta]} \nonumber \\
&\times\Big\{1-\cos[(N+1)\alpha]\cos[(N+1)\beta]\Big\}=0.
 \end{align}
This equation should be solved together with Eq.~\eqref{eq:appA alpha beta1} to give us all
admissible labels. Generically, this has to be done numerically.

In the analysis below, we focus on the regime with $\mu\geq 0$ and $\Delta\geq 0$.
We assume that $\Delta\neq 1$, the case $\Delta = t = 1$ was considered explicitly
in \cite{epw70,p70}. 
From the equations \eqref{eq:appA alpha beta2} and \eqref{eq:appA alpha beta1},
we see that a solution $(\alpha,\beta)$ for $\Delta > 0$ also gives a solution for
$\Delta < 0$ (though the form of the wave function $\phi_{\alpha,n}$ changes).
In addition, the solutions for $\mu <0$ can be related to the solutions with $\mu>0$.
If a pair $(\alpha ,\beta)$ satisfies the equations for $\mu > 0$, the pair $(\alpha +\pi, \beta+\pi)$ will satisfy the equations for
$\mu < 0$. Note that this shift does not change Eq.~\eqref{eq:appA alpha beta2}.
However, it gives rise to a minus sign in the left hand side of Eq.~\eqref{eq:appA alpha beta1} which
indeed changes the sign of $\mu$. Finally, the actual eigenvalues $\Lambda_\alpha$ are also unchanged.

\begin{figure}[t]
\includegraphics[width=\columnwidth]{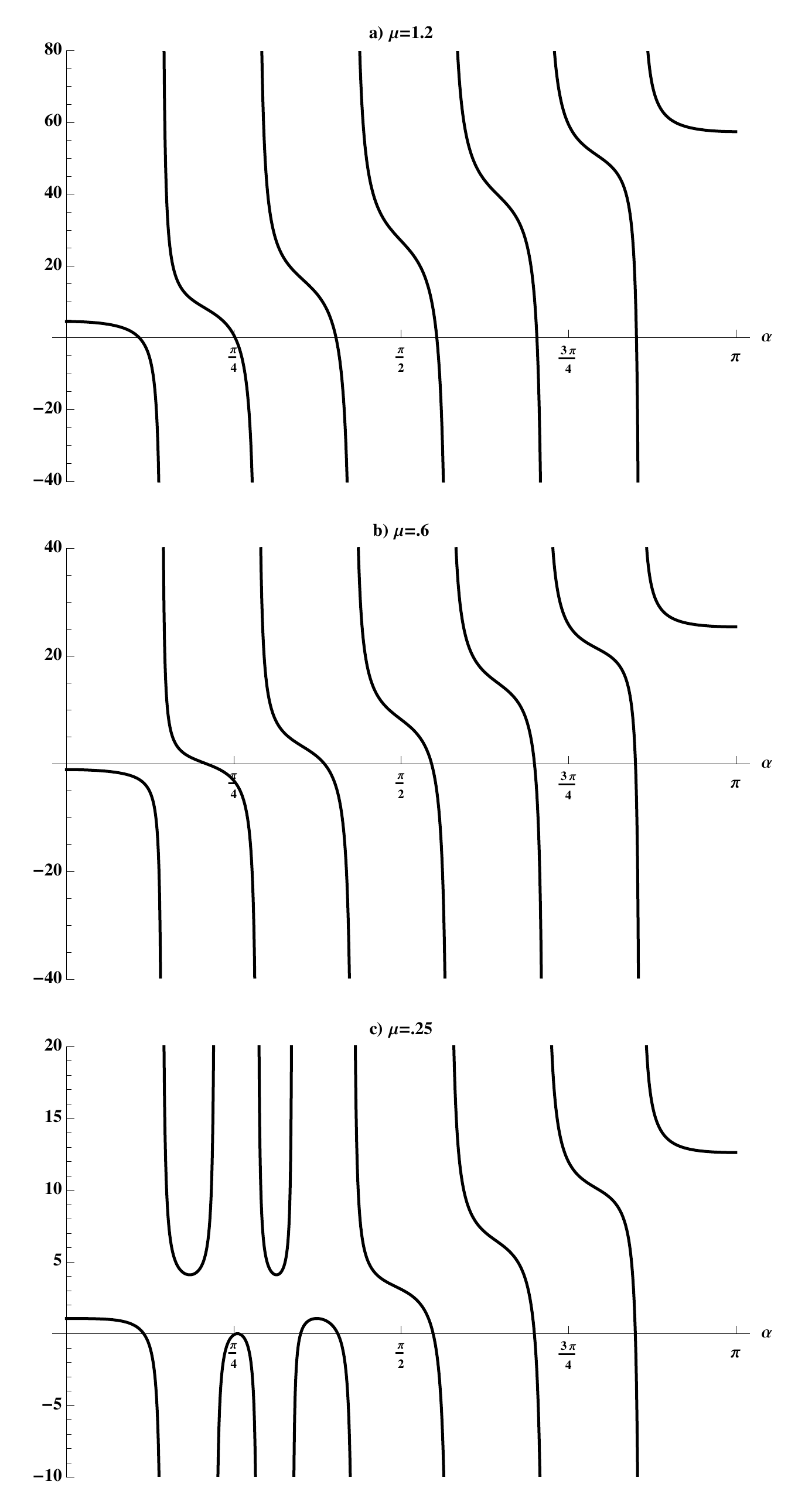}
\caption{
Plot of the left hand side of the constraint Eq.~\eqref{eq:appA alpha beta2} as a function
of $\alpha$ for $N=6$, $\Delta = 0.8$ and $\mu = 1.2, 0.6, 0.25$ for a), b) and c) respectively.
For $\mu = 1.2$, there are six solutions, so there are no zero-modes. For $\mu =  .6$, there
are five solutions. For $\mu = .25$, there are five independent solutions, which one can 
pick to lie in the range $\alpha > \alpha_c \approx 0.25 \pi$. 
}
\label{fig:alpha-plot}
\end{figure}

Thus from now on, we assume that $\mu,\Delta \geq 0$. The structure of the solutions $(\alpha,\beta)$ is as follows.
For $\mu > 1$, one finds $N$ solutions, for which
$\alpha$ and or $\beta$ is real. Because $\alpha$ and $\beta$ are completely equivalent, we assume that
$\alpha$ is real. When $0 \leq \mu < 1$, there are $N-1$ solutions, with $\alpha$ real, and $\beta$ either real
or complex. We note that if $\beta$ is complex, its real part ${\rm Re} \beta = 0$ for $\Delta < 1$, and
${\rm Re} \beta = \pi$ for $\Delta > 1$.
The `missing' solution has both $\alpha$ and $\beta$ complex, and corresponds to the zero mode,
which we describe in detail below. In Fig.~\ref{fig:alpha-plot}, we show this for a chain of $N=6$
sites, $\Delta = 0.8$ and different values of $\mu$.

Before we do so, we first discuss the solutions with $\alpha$ real. 
We first note that for any $(\alpha, \beta)$ pair that solves Eqn.~\eqref{eq:appA alpha beta2}
and \eqref{eq:appA alpha beta1}, all the combinations of $(\pm \alpha,\pm \beta)$ are also a
solution. Since these pairs give rise to same wavefunction, we only consider $\alpha$ in the
range $0 \leq \alpha \leq \pi$.

The solutions are then
obtained by finding the solutions of Eq.~\eqref{eq:appA alpha beta2}, where $\beta$ is given by Eq.~\eqref{eq:appA alpha beta1}.
Special care has to be taken in the case that both $\alpha$ and $\beta$ are real, say $(\alpha,\beta) = (\alpha_1,\beta_1)$,
because one will also find the equivalent solution $(\alpha,\beta) = (\beta_1,\alpha_1)$, so one has to restrict the range of
$\alpha$ further, to avoid `double counting' of solutions.

From Eq.~\eqref{eq:appA alpha beta1}
it is clear that $\alpha$ and $\beta$ can only be both real
when $-1 \leq \frac{\mu}{1-\Delta^2} \leq 1$. Because $\mu,\Delta \geq 0$, this leads to
two regimes, $\Delta < \sqrt{1-\mu}$ and $\Delta > \sqrt{1+\mu}$. In these regimes, one always finds the solution
$\alpha = \beta = \arccos(\mu/(1-\Delta^2)) = \alpha_c$,
because Eq.~\eqref{eq:appA alpha beta2} is trivial when $\alpha = \beta$.
This solution is not valid, however, because it leads to $\phi_{\alpha,n} = 0$.

Nevertheless, the value $\alpha_c$ is useful when specifying the appropriate range
for $\alpha$. If there are solutions with both $\alpha$ and $\beta$ real, one has that
either $\alpha < \alpha_c < \beta$, or $\beta < \alpha_c < \alpha$.
In addition, for the range $\Delta \leq \sqrt{1-\mu}$,
one finds that all the solutions $(\alpha, \beta)$ with $\beta$ imaginary have
$\alpha > \alpha_c$. Thus, to find all solutions in this range, one should only take
the solutions for $\alpha$ such that $\alpha_c < \alpha < \pi$.
For the range $\Delta \geq \sqrt{1+\mu}$, the situation is opposite,
and one should take the solutions for $\alpha$ in the range $0 \leq \alpha < \alpha_c$.
In the other regime, namely $\sqrt{1-\mu} < \Delta < \sqrt{1+\mu}$,
one has to consider all solutions for $\alpha$ in the range $0 \leq \alpha < \pi$.

We now turn our attention to the Majorana zero mode solution.
The goal is to find the analytical expression for the wave function of this mode. For
simplicity, we work in the limit of large system size, i.e., $N\rightarrow\infty$.

By analyzing Eq.~\eqref{eq:appA alpha beta2}, one finds that the solution one loses, is the one
with smallest positive, real $\alpha$.
Taking the limit $\alpha \rightarrow 0$ and $N\rightarrow \infty$ of  
Eq.~\eqref{eq:appA alpha beta2}, using Eq.\eqref{eq:appA alpha beta1}, gives
\begin{equation}
\dfrac{4}{\Delta^2 (1-\Delta^2)} (\mu-1)[\mu - (1-\Delta^2)] = 0
\end{equation}
This shows that there is a solution with $\alpha = 0$, for $\mu=1$. In addition, further analysis shows that
for $\mu < 1$, one loses this solution, both for $\Delta < 1$ and $\Delta > 1$,
while for $\mu > 1$, this solution shifts to finite, positive $\alpha$.
This behavior can be seen for a chain with $N=6$ sites, $\Delta = 0.8$ and $\mu = 1.2, 0.6, 0.25$ 
in Fig.~\ref{fig:alpha-plot}. In the case of $\mu = 0.25$, only the solutions with
$\alpha > \alpha_c \approx 0.25 \pi$ are independent, so the there are still only
five solutions. The additional, sixth solution is still a zero-mode.

We note that for finite $N$, the value of $\mu$ for which one loses the solution has $1/N$
corrections, and depends on $\Delta$. That the phase transition between the trivial and topological phase
occurs for $\mu=1$ in the large $N$ limit is of course well known, and is given by the value of $\mu$ for which
the gap closes. Based on Eq.~\eqref{eq:appA dispersion}, we infer that $\mu=\pm 1$ are the only possible values
of chemical potential for which gap closes (provided that $\Delta \neq 0$).

Now we turn to finding the missing root and its associated features. To do so we need to consider different cases. 

\textit{1) $\Delta <1$ and $\sqrt{1- \Delta^2}<\mu<1$}:
In this regime, we lost one solution with $\alpha$ real, so we look for a solution with both
$\alpha$ and $\beta$ imaginary, and in fact, purely real.
Such a solution indeed exist namely, 
\begin{eqnarray}
\label{eq:appA decay}
&\alpha^*=i(\dfrac{1}{\xi_1} - \dfrac{1}{\xi_2}),&\ \beta^*= i(\dfrac{1}{\xi_1} + \dfrac{1}{\xi_2}),\\
&\cosh \dfrac{1}{\xi_1} = \dfrac{1}{\sqrt{1-\Delta^2}},& \ \cosh \dfrac{1}{\xi_2} = \dfrac{\mu}{\sqrt{1-\Delta^2}} ,
\end{eqnarray}
which solves Eq.~\eqref{eq:appA alpha beta1} and Eq.~\eqref{eq:appA alpha beta2} in the
large $N$ limit.
For $\sqrt{1-\Delta^2} < \mu <1$, both $\xi_1$ and $\xi_2$ are real.
Let us explore the properties of this solution.
First, substituting this result back into the Eq.~\eqref{eq: appA energy level} gives us $\Lambda_{\alpha^*}=0$,
so we indeed have a zero-mode. This means that we can use Eq.~\eqref{eq:phi(A-B)} to solve for the
wave function. 
Alternatively, we can set $A_1=0$ in Eq.~\eqref{eq: appA finalansatz} to obtain the Majorana mode
that is localized on the left side of the system. Either approach gives
\begin{equation}
\label{eq: app A  phi decay}
\phi_{\alpha^*,n}=C e^{-\dfrac{n}{\xi_1}} \sinh(\dfrac{n}{\xi_2}),
\end{equation}
where $C$ is a normalization constant.
Because $\xi_1 < \xi_2$, the mode $\phi_{\alpha^*}$ is indeed localized on the left edge. 

The same reasoning can be done for $\psi_{\alpha,n}$. The important observation is that $(A+B)(A-B)$ has the same structure as $(A-B)(A+B)$ if we look at it from the other side of the chain, i.e. $n \to N+1-n$. So we get $\psi_{\alpha^*,n}=\phi_{\alpha^*,N+1-n}$, which tells us that $\psi_{\alpha^*,n}$ is localized on the right edge.

\textit{2) $\Delta <1$ and $\mu < \sqrt{1- \Delta^2}$}:
For $\mu < \sqrt{1- \Delta^2}$, the parameter $\xi_2$ in Eq.~\eqref{eq:appA decay} becomes imaginary,
so is more natural to rewrite the previous solution.
The root can be written as:
\begin{eqnarray}
\label{eq:appA oscdecay}
&\alpha^* = -q+i\dfrac{1}{\xi},& \ \beta^*= q + i\dfrac{1}{\xi},\\
&\cos q = \dfrac{\mu}{\sqrt{1-\Delta^2}},& \ \cosh \dfrac{1}{\xi} = \dfrac{1}{\sqrt{1-\Delta^2}}. 
\end{eqnarray}

Again, one finds that $\Lambda_{\alpha^*}=0$.
Using the same logic as above, one finds that
\begin{equation}
\label{eq: app A  phi oscdecay}
\phi_{\alpha^*,n}=C e^{-\dfrac{n}{\xi}} \sin(nq),
\end{equation}
with $C$ some constant. This result shows that $\phi_{\alpha^*}$ is localized on the left edge. Although this in this case instead of having decaying functions, we have an oscillatory decaying function.

\textit{3) $\Delta >1$}:
In this case we can not use the previous results, because $\sqrt{1-\Delta^2}$ becomes imaginary. One finds that the new root in this regime is given by
\begin{eqnarray}
\label{eq:appA Delta >1}
&\alpha^*=i(\dfrac{1}{\xi_1}-\dfrac{1}{\xi_2}),&\ \beta^*=\pi + i(\dfrac{1}{\xi_1}+\dfrac{1}{\xi_2}),\\
&\sinh \dfrac{1}{\xi_1} = \dfrac{1}{\sqrt{\Delta^2-1}},& \  \sinh \dfrac{1}{\xi_2} = \dfrac{\mu}{\sqrt{\Delta^2-1}}. 
\end{eqnarray} 
We see that $\xi_1<\xi_2$ since $\mu <1$. One can check that for this root $\Lambda_{\alpha^*}=0$, hence it is also a zero mode. To find the Majorana mode that is localized on the left edge, we again set $A_1 = 0$ in Eq.~\eqref{eq: appA finalansatz}, which results in
\begin{equation}
\phi_{\alpha^*,n}=C e^{-\dfrac{n}{\xi_1}} \times \left\{
\begin{array}{rl}
 \cosh(\dfrac{n}{\xi_2}), & \text{if } n \ \text{is odd} ,\\
 \sinh(\dfrac{n}{\xi_2}) & \text{if } n \ \text {is even}.
\end{array} \right.
\end{equation}
This result shows that $\phi_{\alpha^*}$ is localized on the left edge.

To close this section we note that for $\mu=\sqrt{1- \Delta^2}$, we have $\alpha^*=\beta^*$.
Therefore one can not use $x_{\alpha}^n$ and $x_{\beta}^n$ as separate solutions, but 
one should use $nx_{\alpha}^{n}$ as the other independent solution.


\section{The zero-modes of the Kitaev chain with a phase gradient}
\label{sec:App Kitaev+phase}
In this appendix, we investigate the zero mode of the Kitaev chain, in the
presence of a phase gradient in the order parameter. We assume that $| t | = | \Delta | = 1$.

As we mentioned in Sec.~\ref{sec Pfeutyphase}, after a gauge transformation the Hamiltonian
takes the form
\begin{align}
H&=\dfrac{1}{2}\sum_{j=1}^{N-1}(e^{i \frac{\nabla \theta}{2}}c^{\dagger}_jc_{j+1}+e^{- i \frac{\nabla \theta}{2}}c^{\dagger}_jc^{\dagger}_{j+1}+h.c)\nonumber \\
& -\mu \sum_{j=1}^{N}(c^{\dagger}_jc_{j}-\dfrac{1}{2}),
\end{align}
in which $ \nabla \theta$ is the phase gradient per site, which is constant.
To find the zero-mode, we use the method which is presented in Sec.~\ref{secLSM}.
From Eq.~\eqref{eq:appA gen A-B A+B}, the matrices $A-B$ and $A+B$ read 
\begin{widetext}
\begin{align}
\label{eq:matices pfeuty+ gradient}
&A-B = \left(
\begin{array}{ccccccccc}
-\mu & i \sin(\frac{\nabla \theta}{2}) & & & & &  & &   \\
e^{- i \frac{\nabla \theta}{2}} &-\mu &i \sin(\frac{\nabla \theta}{2}) &  & &  &\mathbf{0} &   \\
0& e^{-i \frac{\nabla \theta}{2}} &-\mu & i \sin (\frac{\nabla \theta}{2})  & &  & &   \\
 & & & &\ddots  & &  & &\\
& & & & & e^{-i \frac{\nabla \theta}{2}}& -\mu &i \sin(\frac{\nabla \theta}{2}) & 0\\ 
 & \mathbf{0}& & & & & e^{-i \frac{\nabla \theta}{2}} &-\mu &i \sin(\nabla \theta\frac{\nabla \theta}{2})\\
 & & & &  & &  & e^{-i\frac{\nabla \theta}{2}}&-\mu 
\end{array} \right),& \\
&A+B =\left(
\begin{array}{cccccccc}
-\mu &\cos(\frac{\nabla \theta}{2}) & & & & &  &    \\
0 &- \mu &\cos(\frac{\nabla \theta}{2}) & &  &   &\mathbf{0} &   \\
 & & & &\ddots  & &   &\\
& & &  & & -\mu & \cos(\frac{\nabla \theta}{2})& 0\\ 
 & \mathbf{0}& & &  &  &-\mu &\cos(\frac{\nabla \theta}{2})\\
 & & & &  &   & &-\mu 
\end{array} \right).&
\end{align}
\end{widetext}
In this case we are only looking for the zero mode and the corresponding Majorana operator.
Hence we drop the $\alpha^*$ index.
In order to have a Hermitian operator, $\psi$ needs to be imaginary in Eq.~\eqref{eq:etapsiphi}. So we set 
$\psi = i \tilde{\psi}$ and $g_n=\dfrac{1}{2}(\phi_n +i \tilde{\psi}_n )$. First we look at the `bulk equation' that
follows from  Eq.~\eqref{eq:LSMimag}:
\begin{equation}
- \mu \tilde{\psi_n} +\cos(\frac{\nabla \theta}{2}) \tilde{\psi}_{n-1}=0.
\end{equation}
The only equation which is different from this bulk equation has index one (note that matrices are acting from the right on the vectors),
\begin{equation}
- \mu \tilde{\psi_1}=0.
\end{equation}
These two equations give us the solution:
\begin{equation}
\label{eq:app B psi}
\tilde{\psi}_n=R \Big[\dfrac{\mu}{\cos(\frac{\nabla \theta}{2})} \Big]^{N-n+1},
\end{equation}
where $R$ is a normalization constant. We see that the boundary equation holds (in the large $N$ limit), provided that
$\mu < \cos(\frac{\nabla\theta}{2})$, which precisely corresponds with the criterion to be in the topological phase, as we
discussed in Sec.~\ref{sec Pfeutyphase}.

We move on to find $\phi$. The `bulk equation' coming from Eq.~\eqref{eq:LSMreal} reads
\begin{equation}
\label{eq:phi-phase-gradient}
-\mu \phi_n + \cos(\frac{\nabla \theta}{2}) \phi _{n+1}=\sin(\frac{\nabla \theta}{2})(\tilde{\psi}_{n-1} - \tilde{\psi}_{n+1}).
\end{equation}
Here we encounter the first difference in comparison with the case with only real couplings.
In this case the equation governing $\phi$ depends on $\tilde{\psi}$.
This means that the general solution for $\phi$ consists of a part that satisfies Eq.~\eqref{eq:phi-phase-gradient} with the
right hand side set to zero, and a particular solution.
The general solution takes the following form
\begin{equation}
\label{eq:app B phi}
\phi_n = L\Big[\dfrac{\mu}{\cos(\frac{\nabla \theta}{2})} \Big]^{n} - R \tan(\frac{\nabla \theta}{2}) \Big[\dfrac{\mu}{\cos(\frac{\nabla \theta}{2})} \Big]^{N-n+1},
\end{equation}
where the first term satisfies Eq.~\eqref{eq:phi-phase-gradient} with the right hand side set to zero
and the second term satisfies the full Eq.~\eqref{eq:phi-phase-gradient}. Thus, in this (unnormalized)
solution, $L$ is a free parameter. We note that the first term is localized
on the left hand side of the system, while the second term is localized on the right hand side.
We should also check the two boundary equations, which are given by:
\begin{eqnarray}
\label{eq:appB boundary1}
&n=1:&  -\mu \phi_1 +\cos (\frac{\nabla \theta}{2})\phi_2= -\sin(\frac{\nabla \theta}{2}) \tilde{\psi}_2, \\
\label{eq:appB boundaryN}
&n=N:& -\mu \phi_N =\sin(\frac{\nabla \theta}{2}) \tilde{\psi}_{N-1}.
\end{eqnarray}
By substituting the solution for $\phi$ and $\psi$ back into Eqs.\eqref{eq:appB boundary1}
and \eqref{eq:appB boundaryN}, we find that
they are satisfied up to terms that are exponentially small in the large $N$ limit.

Using this general solution, we can construct two solutions for $g_n$, that are localized on
either side of the system. Setting $R=0$, one finds a real solution (localized on the left):
$g_n = L \Big[\dfrac{\mu}{\cos(\frac{\nabla \theta}{2})}\Big]^{n}$. Using Eq.~\eqref{eq:etapsiphi},
we see that the corresponding electron operator $\eta_\alpha$ only involves the
operators $\gamma_{A,i}$, not the $\gamma_{B,i}$. The other solution, localized on the
right, is found for $L=0$, and is given by
$g_n = \dfrac{R}{\cos (\frac{\nabla \theta}{2})} i e^{i \frac{\nabla \theta}{2}} \Big[\dfrac{\mu}{\cos(\frac{\nabla \theta}{2})}\Big]^{N-n+1}$.
Thus, this right mode involves both
$\gamma_{A,i}$ and $\gamma_{B,i}$.
We note that the above solutions are valid in the limit of semi-infinite chains. In the case of
a finite, but long chain, they can be combined to form an approximate solution (up to corrections
that are exponentially small the length of the system) of the fermionic zero-mode,
that is delocalized, with support on both ends of the chain.

\end{document}